\documentclass[sigconf]{acmart}

\usepackage{mathbbol}
\newcommand\sig[1]{#1}
\newcommand\crd[1]{#1}

\newcommand{\systemname}{Scones~}
\newcommand{\systemnamenospace}{Scones}
\AtBeginDocument{%
  \providecommand\BibTeX{{%
    \normalfont B\kern-0.5em{\scshape i\kern-0.25em b}\kern-0.8em\TeX}}}

\copyrightyear{2020}
\acmYear{2020}
\setcopyright{rightsretained}
\acmConference[IUI '20]{25th International Conference on Intelligent User Interfaces}{March 17--20, 2020}{Cagliari, Italy}
\acmBooktitle{25th International Conference on Intelligent User Interfaces (IUI '20), March 17--20, 2020, Cagliari, Italy}\acmDOI{10.1145/3377325.3377485}
\acmISBN{978-1-4503-7118-6/20/03}

\begin{document}

\title{Scones: Towards Conversational Authoring of Sketches}


\author{Forrest Huang}
\affiliation{%
  \institution{University of California, Berkeley}
  \city{Berkeley}
  \state{California}
  \country{U.S.A.}}
\email{forrest_huang@berkeley.edu}

\author{Eldon Schoop}
\affiliation{%
  \institution{University of California, Berkeley}
  \city{Berkeley}
  \state{California}
  \country{U.S.A.}}
\email{eschoop@berkeley.edu}

\author{David Ha}
\affiliation{%
  \institution{Google Brain}
  \city{Tokyo}
  \country{Japan}}
\email{hadavid@google.com}

\author{John Canny}
\affiliation{%
  \institution{University of California, Berkeley}
  \city{Berkeley}
  \state{California}
  \country{U.S.A.}}
\email{canny@berkeley.edu}

%\affiliation{\institution{The Th{\o}rv{\"a}ld Group}}
%\email{jsmith@affiliation.org}



% Define the interactions more, maybe inspired from human processes. Conversation -> better 
% doesn't do free text critique yet.
% use __ONE__ way to define the interaction
% collaborative with machine vs. users?

% refer to this iterative process of sketch->text->sketch response as a multimodal conversation?
\begin{abstract}
\crd{Iteratively refining and critiquing sketches are crucial steps} to developing effective designs.
We introduce \systemnamenospace, a mixed-initiative, machine\crd{-}learning\crd{-}driven system \crd{that} enables users to iteratively \crd{author} sketches from text instructions.
\systemname is a novel deep-learning-based system that iteratively generates \emph{scenes} of \emph{sketched objects} composed with semantic specifications from natural language.
\systemname exceeds state-of-the-art performance on a text-based scene modification task, and introduces a mask-conditioned sketching model that can generate sketches with poses specified by high-level scene information.
% what kind of conversation? multimodal?
% novel 
% Scones formulates sketch refinement as a novel task for training machine learning models by sharing sketch state and natural language commands in the same embedding space.
% Scones also 
% To use Scones, users can specify objects to be added to a scene, or modifications (e.g., move and transform) to objects.
% critique to refine a sketch by moving, transforming, and reflecting the objects. Scones achieves state-of-the-art performance for a critiquing task on the CoDraw dataset.
% critiquing -> iterative drawing task? (See codraw paper)
In an exploratory user evaluation of \systemnamenospace, participants reported enjoying an iterative drawing task with \systemnamenospace, and suggested additional features for further applications.
%also consider Scones enjoyable on average in a collaborative sketching task supported by Scones.
We believe \systemname is an early step towards automated, intelligent systems \crd{that} support human-in-the-loop applications for communicating ideas through \crd{sketching} in art and design.
% what specific applications? brainstorming? Iterative drawing? Schematics? Technical drawing? Education?
\end{abstract}




\begin{teaserfigure}
    \centering
  \includegraphics[width=1.0\textwidth]{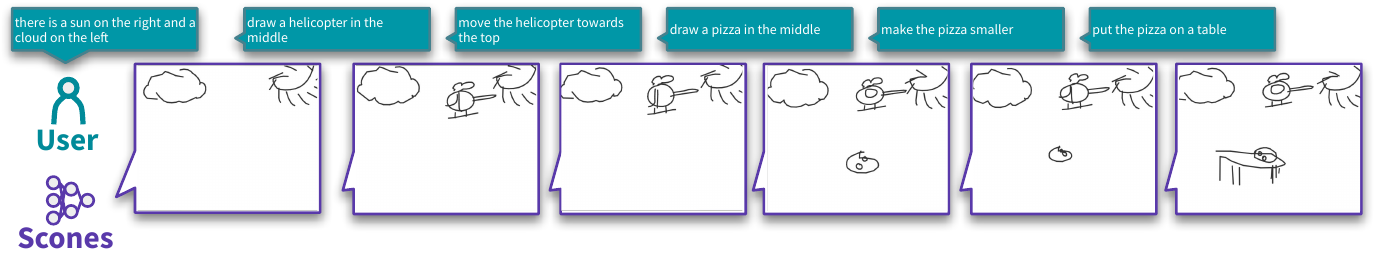}
  \caption{Example interaction between Scones and a human user. Scones can iteratively generate and refine sketched scenes given users' text instructions. }
  \Description[This is a teaser figure with an example interaction between Scones and a human user. The human user supply text instructions for Scones to sketch on the canvas iteratively.]{This is a teaser figure with an example interaction between Scones and a human user. From left to right, the human user takes multiple turns to provide multiple text instructions to Scones. After each instruction, Scones will respond by modifying the sketching canvas by adding or modifying objects. In the beginning, the user tells Scones that `there is a sun on the right and a cloud on the left’, then Scones drew a cloud and a sun on the canvas. Scones subsequently added a helicopter, pizza, and a table to the next few scenes. }
  %\Description[test]{test}
  \label{fig:teaser}
\end{teaserfigure}

\maketitle

\section{Introduction}
Sketching is a powerful communication medium, as even rough drawings can richly communicate the intent of artists, designers, and engineers.
These practitioners use sketches as a tool to iteratively present, critique and refine ideas.
%Many creative processes use sketches as an ideation and communication medium to develop ideas for collaborators to review, critique, and revise.
However, creating sketches that effectively communicate ideas visually requires significant training. Furthermore, the use of sketches in an iterative design process, where the sketch itself is annotated or refined, requires additional, specialized expertise.

%The recent introduction of Machine Learning (ML) models \crd{capable of comprehending and generating} visual and natural language content has increased the potential for intelligent systems to participate in \crd{creative processes}.
%\eldon{
\sig{Recently developed Machine Learning (ML) models have illuminated how intelligent systems can participate in the sketching and critique processes, e.g., by generating sketches for single objects~\cite{sketchrnn}, and using natural language to create images~\cite{semanticMaskLayout}.
However, the broader interaction of iteratively \emph{critiquing} and \emph{refining} complex sketches comprising multiple objects poses several additional challenges.
For this task, a system would need to unify knowledge of the \textit{low-level} mechanics for generating sketch strokes and  natural language modification instructions with a \textit{high-level} understanding of composition and object relationships in scenes.}
%For this task, a system would need to obtain knowledge of the \textit{high-level} composition and relationships of the objects in the sketch and \textit{low-level} mechanics for generating sketch strokes for objects of respective classes, and to understand natural language instructions for modification.}

%While these systems can already assist users in creating photo-realistic images from semantic maps \cite{gaugan} or through text instructions \cite{semanticMaskLayout}, they generally optimize for a single, immutable generated example, or only allow modification through traditional means such as direct manipulation.

In this paper, we introduce \systemnamenospace, an intelligent system for iteratively generating and modifying scenes of \crd{sketched objects} through text instructions. Our contribution is \crd{three-fold:}
\sig{
\begin{itemize}
    \item We formulate the novel interaction of iteratively generating and refining sketches with text instructions and present a web-deployable implementation of  \systemname to support this interaction.
    \item We contribute a scene composition proposer, a component of our system that takes a novel approach in creating and editing scenes of objects using natural language. It adapts a recent neural network architecture and improves state-of-the-art performance on the scene modification task.
    \item We introduce a novel method for specifying high-level scene semantics within individual object sketches by conditioning sketch generation with mask outlines of target sketches.
\end{itemize}
}

\sig{We evaluate our intelligent user interface on an iterative sketching task with 50 participants, where each was asked to use text instructions to create a scene matching a target output. Our results show participants enjoyed the task and were generally satisfied with the output of \systemnamenospace. Participants also provided feedback for improving \systemname in future iterations.}

\sig{Our ultimate goal for \systemname is to support creative processes by facilitating the iterative refinement of complex sketches through natural language. Combining these modalities is a fundamental part of our contributions, as this allows users to freely express their intent using abstract, text-based instructions, together with concrete visual media.}

\section{Related work}

% ES #3 should be copyedited here
Scones builds upon related work in four key areas: (1) deep neural networks that generate sketches and scenes, and corresponding datasets \crd{they were trained on}; (2) sketching support tools that refine and augment sketch inputs; (3) machine-learning-based applications that support image generation from drawing input; and, (4) interfaces that use natural language to interact with visual data.

\subsection{Neural Sketch Generation and Large-scale Sketch Datasets}
Recent advancements in the ML community introduced deep neural networks capable of recognizing and generating sketches. Sketch-RNN~\cite{sketchrnn} is one of the first RNN-based models that can generate sequential sketch strokes through supervised \crd{learning} on sketching datasets.
Generative Adversarial Networks (GANs) have also been used to translate realistic images into sketches (or edges) at the pixel level by training on \crd{paired~\cite{photo-sketching}} and unpaired~\cite{cyclegan} sketch and image data.
While these methods are well-suited for generating sketches of individual objects or stylizing images, they do not encode high-level semantic information of a scene. Sketchforme takes a two-stage approach to generate sketches of scenes comprising multiple objects by first generating a high-level scene layout from text input, and, next, filling the layout with object sketches~\cite{sketchforme}.

These ML techniques rely heavily on large-scale sketching datasets. The Quick, Draw!~\cite{quickdraw} and TU-Berlin~\cite{tuberlin-sketch} datasets consist of human-drawn sketches for 345 and 250 object classes respectively. SketchyDB provides paired images and simple sketches for retrieval tasks~\cite{sketchydb}. \crd{The SketchyScene dataset consists of sketched scenes of pre-drawn objects transformed and resized by humans, as scene sketches are highly demanding for users to create from scratch~\cite{sketchyscenes}.}

Scones builds upon the Sketch-RNN model and Sketchforme's generation process to support progressive, iterative generation and conditional modification of sketched scenes from natural language, a novel machine-learning-driven user interaction. Scones uses a Transformer network~\cite{transformer:vaswani:2017} with a shared natural language and scene information embedding, and is trained on the CoDraw dataset~\cite{codraw} to learn high-level relationships between objects in scenes \crd{and text instructions}.

% ES striking below, put in intro, not here
%In addition, Scones introduces and supports a novel game that provides a mean for users to further create large-scale scene sketch datasets of raw strokes corresponding to multi-turn text descriptions in conversation.

\subsection{Interactive Sketching Tools}
Research in the \crd{Human-Computer Interaction (HCI) community} has produced interfaces that use drawing input for creating interactive media and prototypes. SILK enables users to annotate user interface mockup sketches to create interactive prototypes~\cite{silk}. Other work adapts these metaphors for creating interactive, animated images from drawings~\cite{kitty}. Closely related to our domain, DrawAFriend uses crowdsourced data through a Game With a Purpose (GWAP)~\cite{gwap} to refine and correct users' sketch strokes~\cite{drawingCrowdsourcing}. PixelTone additionally uses natural language speech input to apply filters to annotated images~\cite{PixelTone:Laput}. \sig{Most relevant to \systemnamenospace, Ribeiro and Igarashi introduced a two-way sketch-based communication game for users to iteratively edit sketches using a direct manipulation interface~\cite{sketch-editing}}. \systemname uses natural language input to create and modify sketches, rather than requiring direct pen stroke input.

% Other tools use crowdsourcing to assist users in turning rough specifications into higher fidelity prototypes. Apparition and SketchExpress allow designers to rapidly prototype user interfaces using rough sketches and verbal instructions which crowd workers turn into functional UI elements and animations~\cite{apparition, sketchexpress}.
% Scones draws upon similar metaphors to these tools, where a user provides critique (e.g., ``make the tree smaller'') to Scones and evaluates its output. We also see opportunities to use Scones in a GWAP formulation to collect crowd-generated critiques of sketched scenes.

% ES killing the HCI content for community critique -- only crowdcdrit, almostanexpert relevant only to GWAP in future work
%The process of critiquing and revision is critical to producing successful designs~\cite{tohidi, crowdCrit}. HCI researchers have introduced interfaces enabling remote and crowd workers to provide design critique across multiple domains, including video production~\cite{vidcrit} and visual art~\cite{mosaic:critique}.  Several interfaces solicit design critique from crowds by decomposing this activity into subtasks, such as CrowdCrit~\cite{crowdCrit, almostAnExpert} and CommunityCrit~\cite{communitycrit}.

\subsection{Interactive Image Generation}

Researchers have also explored interactive image generation, particularly for GAN-based methods. iGAN fills user-provided outlines with generated image textures~\cite{igan:zhu:2016}. More recent work has enabled finer-grained control of the output by providing tools for \crd{drawing semantic maps} for generating artwork~\cite{2bit} and photorealistic images~\cite{gaugan}. These approaches have also been extended to fill users' drawn outlines with image textures~\cite{sketchnfill:ghosh2019} and to generate realistic clothing items in a user-specific recommender system~\cite{vizfashiongan}.
While these methods allow users to control the content of and iteratively add to generated images, they rely on a \crd{direct} \emph{visual-to-visual} mapping \crd{between} input and output media to transfer user intent to the canvas. In contrast, Scones uses a \emph{language-to-visual} mapping, enabling users to add to and modify sketches using natural language, a higher-level medium that allows for variation within user specifications. 

%The sketching domain that \systemname targets also has distinct technical and usability features, such that sequences of sketching strokes generated by \systemname \crd{naturally support} iterative modification by users. 
%\eldon{Last sentence here seems a bit pie-in-the-sky. What domain? How are these features more "natural" -- couldn't a visual-to-visual mapping be the most natural? I'd suggest killing it.}

%The perception of sketches and images are distinct tasks~\cite{hype:zhou}, as sketches may have varying fidelity and the end goal may be only to quickly communicate intent.

%Though some ML architectures support using text input to propose layouts in a scene to be filled with image textures, facilitating a \emph{language-to-visual} mapping~\cite{semanticMaskLayout}, 

\subsection{Interfaces Supporting Natural Language Interactions with Visual Data}

Several novel user interfaces and ML models have been developed to use natural language input in interactive visual tasks, \crd{with} a \emph{language-to-visual} mapping. An example ML challenge in this domain is Visual Question Answering (VQA), where a model is provided \crd{with} a target image and a natural language question as input, and outputs a response predicated on visual, lingual, and commonsense knowledge~\cite{vqa:Agrawal:2017}. Other challenges extend this by requiring \emph{justification} of the response~\cite{zellers2019vcr, clevr:cvpr}, or the truthfulness of an input statement relating two images~\cite{nvlr:2017:acl}. These questions in these challenges can be answered by ML architectures such as Relation Networks (RNs), which infer object relationships from the outputs of RNNs and Convolutional Neural Networks (CNNs)~\cite{nnRelationalReasoning:santoro:nips}. Alternatively, VizWiz \crd{deploys} just-in-time crowdsourcing tasks to answer open-ended questions about an image for visually impaired users~\cite{vizwiz:Bigham:2010}.

%DH: truthfulness of an input statement sounds better

%Search and recommendation systems can let users find relevant images from natural langauge queries. 
Deep learning models have been used for image retrieval from natural language captions~\cite{visualsemanticembed}, and algorithmic approaches have been used for searching within videos~\cite{sceneskim}. Adaptive interfaces can also draw correspondence between language and visual block manipulation tasks~\cite{sidaLanguageGames}. Fashion interfaces which recommend items from natural language specifications~\cite{whatigonwear:nlfashion} or by connecting users to stylists through a chatbot~\cite{futureFashion:ranjitha} require knowledge of items' semantic and visual features, as well as highly variant user preferences.
%is a popular domain, as recommending clothing items requires knowledge of the article's semantic and visual features, as well as highly variant user preferences~\cite{fashionrec, deepstyle}. Fashion recommender interfaces have used data-driven approaches to recommend items from natural language input~\cite{whatigonwear:nlfashion}, and chatbots which connect users with trained stylists~\cite{futureFashion:ranjitha}.

\systemname also uses a \emph{language-to-visual} mapping, adapting a state-of-the-art deep neural network \crd{to use} natural language input for a novel visual output task: interactive sketch creation and modification.

% Recent research work have focused on developing conversational systems driven by deep neural networks on various tasks. Researchers have adopted advanced language models, such as GPT-2 and memory networks, and large-scale datasets to train these networks to perform various tasks. These tasks range from question-answering on fundamental knowledge of natural images, to providing fashion recommendations based on user preferences and scheduling events based on users' agenda and itineraries. \todo{replace with real work}

% DH: GPT-2 is not on arxiv, search "Language models are unsupervised multitask learners" on google scholar

% The system architecture of Scones is novel such that it handles user inputs and sketch generation outputs that spans across the domains of natural language and abstract graphics using a single, unified network, in contrast to prior work that relies on separate encoder and decoder networks.

% Justify System Architecture [done?]
% have mask in 3.2 [done]
% section 4 can be shorter, to get section 5.
% Presentation for table 1

\section{System Architecture}
\begin{figure}[h]
  \centering
  \includegraphics[width=\linewidth]{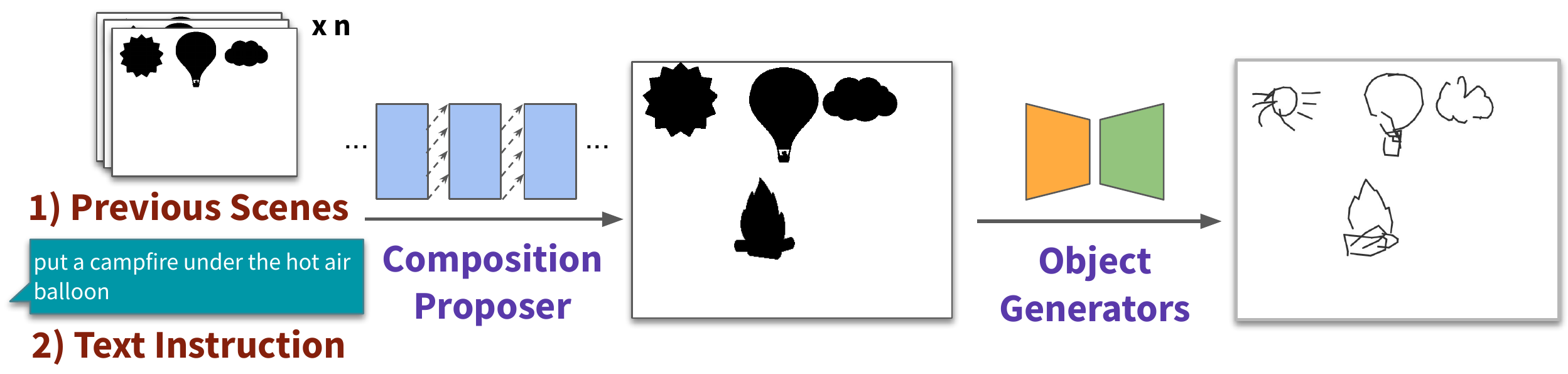}
  \Description[This figure shows the overall architecture of Scones. Scones generates sketched scenes first with the Composition Proposer, then with the Object Generators.]{This figure shows the overall architecture of Scones. We show the two inputs that Scones uses on the left, which are previous scenes and a text instruction. Scones pass these input through the Composition Proposer, which generates binary masks of objects on the canvas. Scones subsequently draws each of these objects using its Object Generators on the right.}
  \caption{Overall Architecture of Scones. Scones takes a two-stage approach towards generating and modifying sketched scenes based on users' instructions.}
  \label{fig:sys}
\end{figure}

The creation of complex sketches often begins with semantic planning of scene objects. Sketchers often construct high-level scene layouts before filling in low-level details.
Modeling ML systems after this high-to-low-level workflow has been beneficial for transfer learning from other visual domains and for \crd{supporting} interactive interfaces for human users~\cite{sketchforme}.
% This workflow have also shown to be beneficial to machine-learning-driven sketching systems for transfer learning from other related visual domains and for increased interactivity between ML models and human users \cite{sketchforme}.
Inspired by this high-to-low-level process, \systemname adopts a hierarchical workflow that first proposes a \crd{scene-level} composition \crd{layout} of objects using its \emph{Composition Proposer}, then generates individual object sketches, conditioned on the scene-level information, using its \emph{Object Generators} (Figure~\ref{fig:sys}).
% considers the relations of multiple objects to user-specified critiques using its \emph{Composition Proposer}, then sketch out individual scene objects using its \emph{Object Generators} (Figure~\ref{fig:sys}). 

\subsection{Composition Proposer}
The Composition Proposer in \systemname uses text instructions to place and configure objects in the scene. It also considers recent past iterations of text instructions and scene \emph{context} at each conversation turn. As text instructions and sketch components occur sequentially in time, each with a variable length of tokens and objects, respectively, we formulate composition proposal as a sequence modeling task. We use a decoder-only Transformer model architecture similar to GPT-2~\cite{gpt2}, a recent deep-learning-based model with high performance.
% As text instructions and sketch modifications occur sequentially in time, and that there are variable number of words in the text critique and variable number of objects in the scenes at various turns that could both be arranged into sequences, in the Composition Proposer we formulate the composition modification task as a sequence modeling task and use a decoder-only Transformer architecture similar to \cite{gpt2}, a recent deep-learning-based sequence model with high performance on sequence modeling tasks such as text generation.

To produce the output scene $S_i$ at turn $i$, the Composition Proposer takes inputs of $n = 10$ previous scenes $S_{(i - n), \dots ,(i - 1)}$ and text instructions $C_{(i-n), \dots , (i - 1)}$ as recent \emph{context} of the conversation. Each output scene $S_i$ contains $l_i$ objects $o_{(i, 1), \dots , (i, l_i)} \in S_i $ and special tokens $o_s$ marking the beginning and $o_e$ marking the end of the scene. Each text instruction $C_i$ contains $m_i$ text tokens $t_{(i, 1), \dots , (i, m_i)} \in C_i$ that consists of words and punctuation marks. 

We represent each object $o$ as a 102-dimensional vector $o = [\mathbb{1}_{s}, \mathbb{1}_{e}, e^{(o)}, e^{(u)}, e^{(s)}, e^{(f)}, x, y]$. The first two dimensions $\mathbb{1}_{s}, \mathbb{1}_{e}$ are Boolean attributes reserved for the start and end of the scene object sequences. $e^{(o)}$ is a 58-dimensional one-hot vector\sig{\footnote{\sig{an encoding of class information that is an array of bits where only the corresponding position for the class to be encoded is 1, and all other bits are 0s.}}} representing one of 58 classes of the scene objects. $e^{(u)}$ is a 35-dimensional one-hot vector representing one of 35 sub-types (minor variants) of the scene objects. $e^{(s)}$ is a three-dimensional one-hot vector representing one of three sizes of the scene objects. $e^{(f)}$ is a two-dimensional one-hot vector representing the horizontal orientation of whether the object is flipped in the x-direction. The last two dimensions $x, y \in [0, 1]$ represents the x and y position of \sig{the center of the object}. This representation is very similar to that of the CoDraw dataset the model was trained on, described in detail in Section \ref{sec:codraw}. For each text token $t$, we use a 300-dimensional GLoVe vector trained on 42B tokens from the Common Crawl dataset~\cite{glove} to semantically represent these words \crd{in} the instructions.

To train the Transformer network with the \crd{heterogeneous} inputs of $o$ and $t$ across the two modalities, we create a unified representation of cardinality $|o| + |t| = 402$  and adopt $o$ and $t$ to this representation by simply padding additional dimensions in the representations with zeros as shown in Equation \ref{eqn:padding1}.

\begin{equation}
\label{eqn:padding1}
    o'_{i, j} = [o_{i, j}, \Vec{0}_{(300)}] \quad
    t'_{i, j} = [\Vec{0}_{(102)}, t_{i, j}]
\end{equation}
 
\begin{figure}[h]
  \centering
  \includegraphics[width=\linewidth]{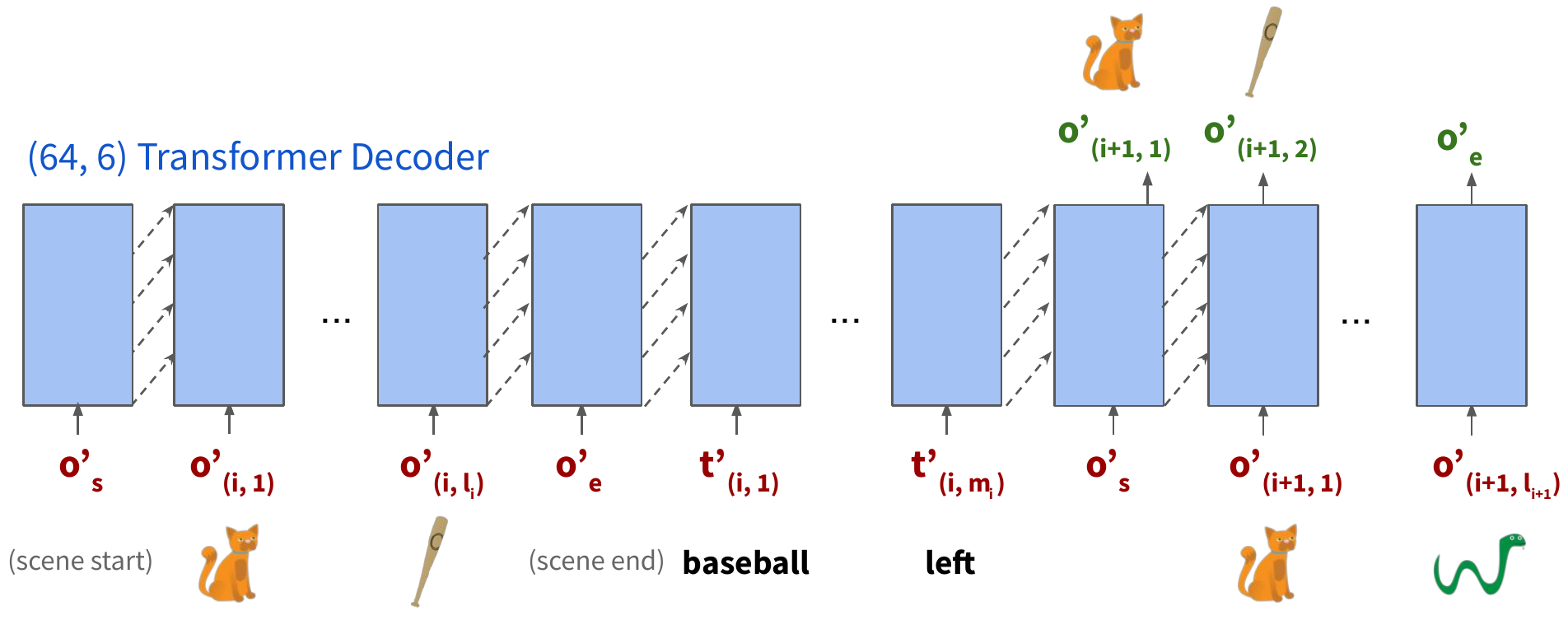}
  \caption{The Scene Layout Generation Process using the Transformer Model of the \textit{Composition Proposer.}}
  \Description[This figure shows the generation process of each scene object by the Transformer model of the Composition Proposer.]{This figure shows the generation process of each scene object by the Transformer model of the Composition Proposer. For each scene, we pass in a start token, the object representation, the end token, the text tokens of the instructions respectively, then generate the final scene sequentially by passing in the current generated object to obtain the next object.}
  \label{fig:transformer}
\end{figure}

We interleave text instructions and scene objects chronologically to form a long sequence $[C_{(i-n)}, S_{(i-n)}$ $, ... , C_{(i-1)}, S_{(i-1)}, C_i]$ as input to the model for generating an output scene representation $S_{i}$. We expand the sequential elements within $C$ and $S$, and add separators to them to obtain the full input sequence to \crd{a single} Transformer Decoder. To adapt the Transformer model to our multi-modal inputs $t'$ and $o'$ and produce new scene objects $o$, we employ a 402-dimensional input embedding layer and 102-dimensional output embedding layer \crd{in the Transformer model}. The outputs from the network are then passed to sigmoid and softmax activations for object position and other properties respectively. We show this generation process in Equation \ref{eqn:transformer} and in Figure \ref{fig:transformer}.

%To our knowledge, our Scene Modifier is the first model to utilize a single model to handle inputs and outputs across multiple modalities, contrary to encoder-decoder architectures commonly used for multi-modality and multi-domain tasks. 

\begin{multline}
\label{eqn:transformer}
    S_i = [o_{(i, 1), ... , (i, l)}] = 
    \textbf{Transformer}([o'_s, o'_{(i - n, 1)}, ... o'_{(i - n, l_{(i-n)})}, \\ o'_e, t'_{(i - n, 1)}, ... , t'_{(i - n, m_{(i - n)})}, ... , t'_{(i, 1)}, ... t'_{(i, l_i)}, o'_s])
\end{multline}

\label{sec:object-sketchers}
\subsection{Object Generators}
Since the outputs of the Composition Proposer are \crd{scene layouts consisting of} high-level object \crd{specifications}, we generate the final raw sketch strokes for each of these objects based on their \crd{specifications} with \emph{Object Generators}. We adopt Sketch-RNN to generate sketches of individual object classes to present to users for evaluation and revision in the next conversation turn. Each sketched object $Q$ consists of $h$ strokes $q_{1 \dots h}$. The strokes are encoded using the \emph{Stroke-5} format~\cite{sketchrnn}. Each stroke $q = [\Delta x, \Delta y, p_d, p_u, p_e]$ represents states of a pen performing the sketching process. The first two properties $\Delta x$ and $\Delta y$ are offsets from the previous point that the pen moves from. The last three elements $[p_d, p_u, p_e]$ are a one-hot vector representing the state of the pen after the current point (pen down, pen up, end of sketch, respectively).
%$p_d = 1$ represents the pen is touching the paper after this point, $p_u = 1$ represents the pen is removed from the paper after this point, $p_e = 1$ represents the sketch has ended.
All sketches begin with the initial stroke $q_1 = [0, 0, 1, 0, 0]$.

%While sizes and positions of each objects in the scene can be trivially addressed by scaling and adding offsets to all strokes in the sketches, and the class of the output sketch can be controlled by employing models trained on sketches of different classes, 

Since Sketch-RNN does not constrain aspect ratio\crd{s}, direction\crd{s} and pose\crd{s} of its output sketches, we introduce two additional conditions for the sketch generation process: masks $m$ and aspect ratios $r$. These conditions ensure our Object Generators generate sketches with appearances that follow the object specifications generated by the Composition Proposer. For each object \crd{sketch}, we compute the aspect ratio $r = \dfrac{\Delta y}{\Delta x}$ by taking the distance between the leftmost and rightmost stroke as $\Delta x$ and the distance between topmost and bottommost stroke as $\Delta y$.
To compute the object mask $m$, we first render the strokes into a pixel bitmap, then mark all pixels as 1 if they are in between the leftmost pixel $py_{xmin}$ and rightmost pixel $py_{xmax}$ that \crd{are passed through by} any strokes for each row $y$, or if they are in between the bottommost pixel $px_{ymin}$ and topmost pixel $px_{ymax}$ that \crd{are passed through by} any strokes for each column $x$ (Equation~\ref{eqn:mask}). As this mask-building algorithm only involves pixel computations, we can use the same method to build masks for clip art objects (used to train the Composition Proposer) to generate sketches with poses matching the Composition Proposer's object representations.

\begin{equation}
\label{eqn:mask}
    m_{(x, y)} =
\left\{
	\begin{array}{ll}
		1  & \mbox{if } py_{xmax} \geq x \geq py_{xmin}, \mbox{or}; \\
		1 &\mbox{if } px_{ymax} \geq y \geq px_{ymin} \\
		0 & \mbox{otherwise}
	\end{array}
\right.
\end{equation}

\begin{figure}[h]
  \centering
  \includegraphics[width=\linewidth]{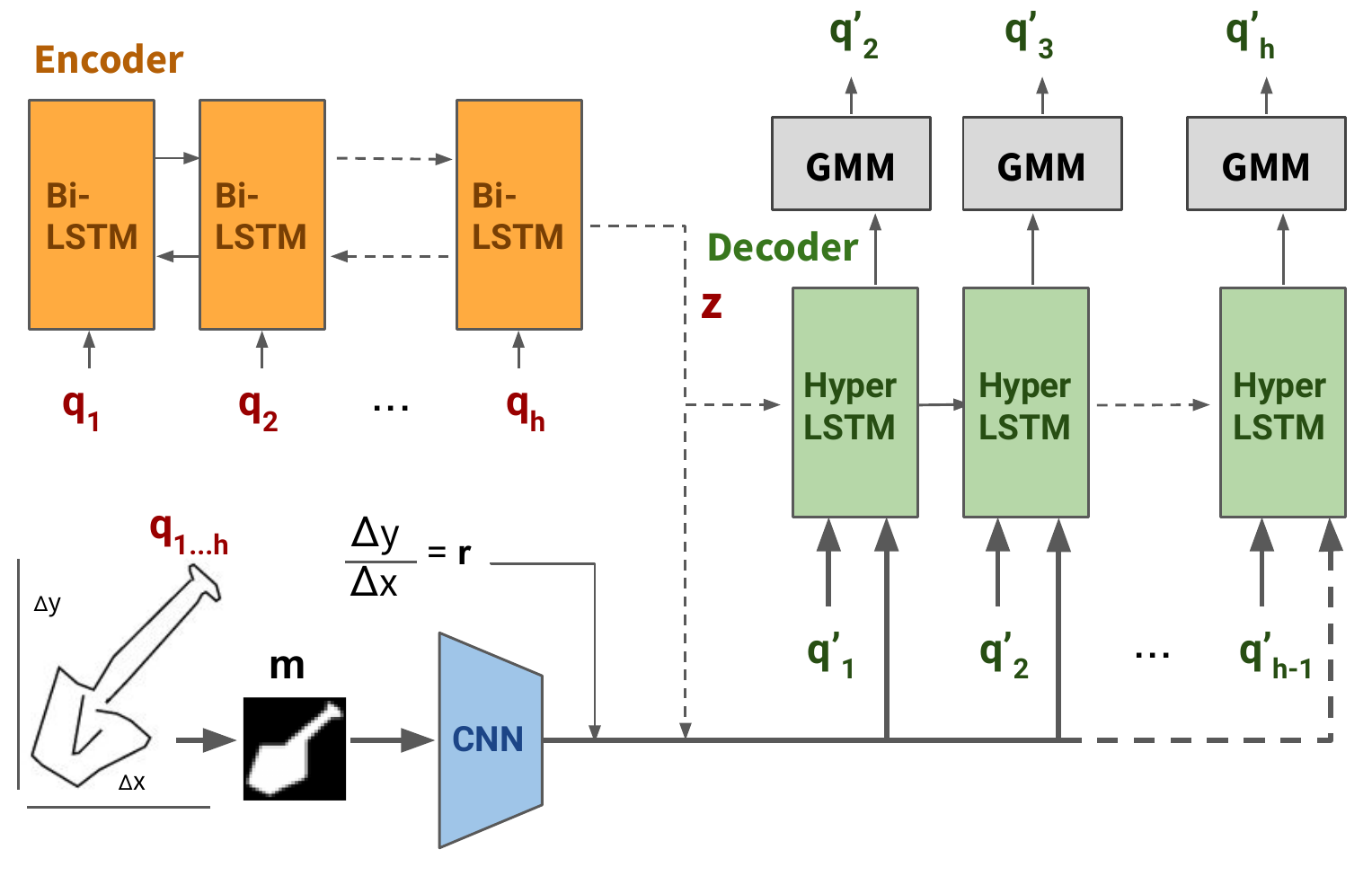}
  \caption{Sketch-RNN Model Architecture of the Object Generators.}
  \Description[This figure shows the Sketch-RNN model architecture of the Object Generators.]{This figure shows the Sketch-RNN model architecture of the Object Generators. On the top left, we have a Bi-LSTM encoder that encodes sketch strokes. On the right, we have a HyperLSTM decoder that decodes each stroke sequentially into Gaussian Mixture Model parameters. These two components largely follow the original Sketch-RNN architecture. In addition to the original Sketch-RNN architecture, in the bottom left we compute the aspect ratio and the mask from the original sketch, and pass the mask through a CNN as additional inputs to the stroke generating Sketch-RNN model.
}
  \label{fig:sketch-rnn}
\end{figure}

We adopt the Variational-Autoencoder(VAE)-based conditional variant of Sketch-RNN to enable generating and editing of sketch objects. Our adopted conditional Sketch-RNN encodes input sketches with a Bi-directional LSTM to a latent vector $z$. The Hyper-LSTM decoder then recreates sketch strokes $q'_{1 \dots h}$ from $z$, and $m, r$ described above during training, as defined in Equation \ref{eqn:sketch-rnn} and shown in Figure \ref{fig:sketch-rnn}. Since the latent space is also trained to match a multi-variate Gaussian distribution, the Object Generator can support sketch generation when the objects are first added to the scene by randomly sampling $z \sim N(0, 1)^{128}$.

\begin{multline}
\label{eqn:sketch-rnn} 
q'_{1 \dots h} = \textbf{Sketch-RNN Decoder}([m, r, z]), z \sim N(0, 1)^{128} \\
z = \textbf{Sketch-RNN Encoder}(q_{1 \dots h})
\end{multline}

As $m$ is a two-dimensional mask, we encode $m$ using a small convolutional neural network into a flattened embedding to be concatenated with $z$, $r$ and $q_i$ as inputs to the decoder. The decoder then outputs parameters for a Gaussian Mixture Model (GMM) which will be sampled to obtain $\Delta x$ and $\Delta y$. It also outputs probabilities for a categorical distribution that will be sampled to obtain $p_d, p_u$ and $p_e$. This generation process and the architecture of the model are illustrated in Figure \ref{fig:sketch-rnn}, and are described in the Sketch-RNN paper~\cite{sketchrnn}.

\section{Datasets and Model Training}
As \systemname uses two components to generate scenes of sketched objects, it is trained on two datasets that correspond to the tasks these components perform.
% As Scones take a two step process towards generating sketches in the critique process using two components, it is trained on two datasets that correspond to each of the tasks that these components perform.

\subsection{CoDraw Dataset}
\label{sec:codraw}
We used the CoDraw dataset to train the Composition Proposer to generate high-level scene \crd{layout} proposals from text instructions. The task used to collect this data involves two human users taking on the roles of \emph{Drawer} and \emph{Teller} in each session. First, the Teller is presented with an abstract scene containing multiple \crd{clip art} objects in certain configurations, and the Drawer is given a blank canvas. The Teller provides instructions using only text in a chat interface to instruct the Drawer \crd{on} how to modify clip art \crd{objects} in the scene. The Teller has no access to the Drawer's canvas in most conversation turns, except in one of the turns when they can decide to `peek' at the Drawer's canvas. 
The dataset consists of 9993 sessions of conversation records, scene modifications, and ground-truth scenes.  
%As briefly described in the encoding of the Composition Proposer, all objects in the scenes belong to one of 58 classes. 

Using this dataset, we trained the Composition Proposer to respond to user\crd{s'} instructions given past instructions and scenes. We used the same training/validation/test split \crd{as} the original dataset.
%We use a Transformer Decoder of size (64, 6) with a 402-dimensional input embedding layer and a 102-dimensional output embedding layer.
Our model is trained to optimize the loss function $L_{cm}$ that corresponds to various attributes of the scene objects in the training set: 

\begin{equation}
L_{cm} = L_{c} + \lambda_{\textbf{sub}} L_{\textbf{sub}} + \lambda_{\textbf{flip}} L_{\textbf{flip}} + \lambda_{\textbf{size}} L_{\textbf{size}} + \lambda_{xy} L_{xy}    
\end{equation}

$L_{c}$ is the cross-entropy loss between the one-hot vector of the true class label and the predicted output probabilities by the model. Similarly $L_{\textbf{flip}}$ and $L_{\textbf{size}}$ are cross-entropy losses for the horizontal orientation and size of the object. $L_{xy}$ is the Euclidean Distance between predicted position and true position of the scene object. We trained the model using an Adam Optimizer with the learning rate of $lr = 1 \times 10^{-4}$ for 200 epochs. We set $\lambda_{\textbf{sub}} = 5.0 \times 10^{-2}$, $\lambda_{\textbf{flip}} = 5.0 \times 10^{-2}$, $\lambda_{\textbf{size}} = 5.0 \times 10^{-2}$, $\lambda_{xy} = 1.0$. These hyper-parameters were tuned based on empirical experiments on the validation split of the dataset. 

\subsection{Quick, Draw! Dataset}
The Quick, Draw! dataset consists of sketch strokes of 345 concept categories created by human users in a game in 20 seconds. We trained our \crd{34} Object Generators on 34 categories of Quick, Draw! data to create sketches of individual sketched objects.

Each sketch stroke in Quick, Draw! was first converted to the Stroke-5 format. $\Delta x$s and $\Delta y$s of the sketch strokes were normalized with their standard deviations for all sketches in their respective categories. Each category consists of 75000/2500/2500 sketches in the training/validation/test set. 

The loss function of the conditional Sketch-RNN $L_{s}$ consists of the reconstruction loss $L_R$ and KL loss $L_{KL}$:

\begin{equation}
    L_s = \lambda_{KL} L_{KL} + L_R
\end{equation}

The KL loss $L_{KL}$ is the KL divergence between the encoded $z$ from the encoder and $N(0, 1)^{128}$. The reconstruction loss $L_R$ is the negative log-likelihood of the strokes under the GMM and a categorical distribution parametrized by the model. We refer interested readers to a detailed description of $L_{s}$ in the original Sketch-RNN paper~\cite{sketchrnn}.
%The RNN model used in Object Generators are 2048-unit HyperLSTM cells.
The initial learning rate of the training procedure was $lr = 1.0 \times 10^{-3}$ and exponentially decayed to $1.0 \times 10^{-5}$ at a rate of $0.9999$. $\lambda_{KL}$ was initially $0.01$ and exponentially increased to $0.5$ at a rate of $0.99995$. The models were also trained with gradient clipping of $1.0$.

\section{Results}
To compare the effectiveness of \systemname at generating scene sketches with existing models and human-level performance, we quantitatively evaluated its performance in an iterative scene authoring task. Moreover, as \systemname uses generative models to produce object sketches, we qualitatively evaluated a large number of examples generated by various stages in \systemnamenospace.
% We evaluated \systemname quantitatively to tangibly demonstrate the effectiveness of it compared to existing models/humans. Moreover, as Scones is a generative model, we further included a large number of examples of some subtasks and complete workflows to illustrate it's performance qualitatively. 

% No Quick, Draw here, state this. Clip-art
\subsection{Composition Modification State-of-the-art}
To evaluate the output of the Composition Proposer against the models introduced with the CoDraw dataset, we adapted its output to match that expected by the well-defined evaluation metrics \crd{proposed by} the original paper~\cite{codraw}.
The original task described in the CoDraw paper involves only proposing and modifying high-level object representations in scenes agnostic to their appearance.
The performance of a ``Drawer'' (a human or machine which generates a scene composition) can be quantified by a similarity metric \sig{constrained between 0 and 5} (higher is more similar) \crd{by comparing properties of and relations between objects in the generated scene and objects in the ground truth from the dataset.}

%The similarity metric consists of a unary component, which compares the properties of individual objects in the scenes, and a pairwise component, which compares the relations between pairs of objects in the scenes. 
Running our Composition Proposer on the CoDraw test set, we achieved an average similarity metric of $3.55$. This exceeded existing state-of-the-art performance (Table~\ref{tab:sota}) on the iterative scene authoring task using replayed text instructions from CoDraw.

%We run the Composition Proposer and obtained the created scene by our model for each interactive sessions in the test set, and achieved an average similarity metric of $3.55$. This is state-of-the-art performance on the task when comparing against other models that uses replayed session sciprts as Tellers in prior work shown in Table \ref{tab:sota}, as reported by CoDraw.

%The output of the Composition Proposer can be reformatted to compare against the models introduced with the CoDraw dataset. As such, we can objectively compare the Composition Proposer's performance with the original models with well-defined quantitative metrics in the original paper.
%The original critqiue task described in the CoDraw paper involves only modifying the object representations of scenes agnostic to sketch strokes used to depict objects in \systemname{}.
%The performance of Drawer(either human or machines)'s performance can be quantified using a five-point similarity metric (higher is more similar) by comparing the final output of the drawer with the ground-truth dataset.

\begin{table}[h]
  \caption{Performance of Various Models on CoDraw Task}
  \Description[This table lists the performance of various models on the CoDraw Task.]{This table lists the performance of various models on the CoDraw Task. Each row describes a teller, a drawer, and the similarity metric value from zero to five. All rows have the script as the teller. Scones achieves a similarity metric of 3.55 at the top row. Neural Network presented in the CoDraw paper achieves 3.39 in the second row. Nearest-Neighbor drawer presented in the CoDraw paper achieves 0.94 in the third row. Human drawers achieve 3.83 in the last row.}
  \label{tab:sota}
  \begin{tabular}{ccc}
    \toprule
    Teller & Drawer & Similarity $\uparrow{}$ (out of 5)\\
    \midrule
    Script & \textbf{\systemname{}} & \textbf{3.55}\\
    Script &  Neural Network~\cite{codraw} & 3.39\\
    Script & Nearest-Neighbour~\cite{codraw} & 0.94 \\
    \midrule
    Script & Human & \textbf{3.83}\\
  \bottomrule
\end{tabular}
\end{table}

\begin{figure}[h]
  \centering
  \includegraphics[width=\linewidth]{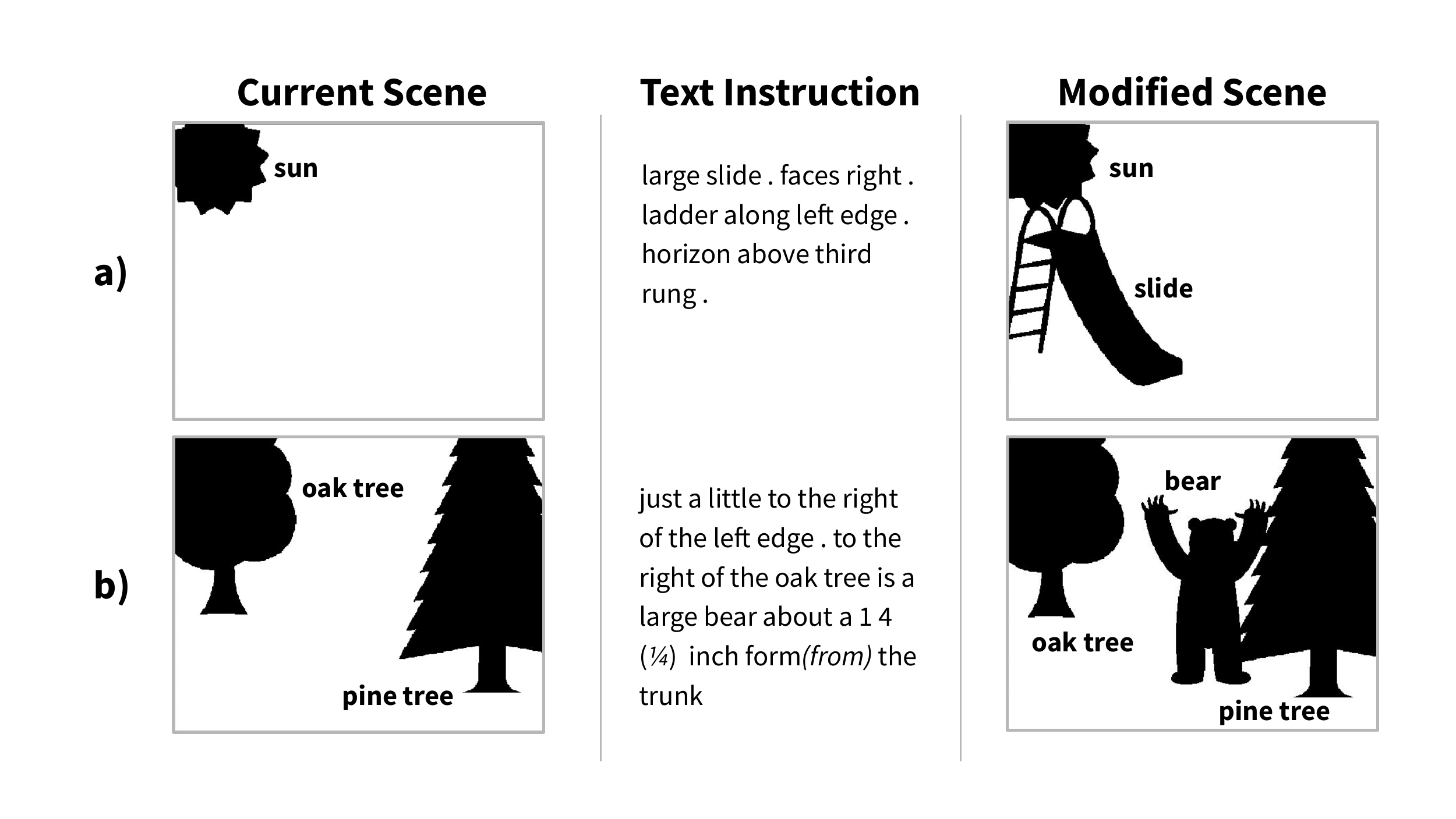}
  \caption{Example Scenes for the Scene Layout Modification Task. The Composition Proposer was able to achieve state-of-the-art performance for modifying object representations in scene compositions.}
  \Description[This figure shows two example scenes for the composition proposal task.]{This figure shows two example scenes for the composition proposal task. Each scene has an original `current scene' on the left, a text instruction in the middle, and a modified scene on the right. Scene a is at the top, and scene b is at the bottom. Scene a originally has only a sun object on the top left. The human instructs the Composition Proposer with ``large slide. faces right . ladder along left edge. horizon above third rung .'' The Composition Proposer adds a slide right below the sun. In scene b, the original scene has an oak tree on the top left, and a pine tree on the bottom right. The human instructs the system ``just a little to the right of the left edge . to the right of the oak tree is a large bear about a fourth inch from trunk''. The system responds by adding a bear in between the pine tree and the oak tree.}
  \label{fig:composition}
\end{figure}

To provide an illustrative example of our Composition Proposer's output on this task, we visualize two example scenes generated from the CoDraw validation set in Figure~\ref{fig:composition}. 
%We further demonstrate the format of the task and the performance of the Composition Proposer qualitatively in Figure~\ref{fig:composition}.
%The Composition Proposer is able to recognize concept classes and object properties from text instructions, and place objects at adequate positions in the scene.
In the \crd{scene a)}, the Composition Proposer extracted the class (slide), direction (faces right), and position relative to parts of the object (ladder along left edge) \crd{from the text instruction}, to place a slide in the scene.
Similarly, it was able to place the bear in between the oak and pine trees \crd{in scene b)}, with the bear touching the left edge of the pine tree.
It is important to note the Composition Proposer completely regenerates the entire scene at each conversation turn. This means it correctly preserved object attributes from previous scenes while making the requested modifications from the current turn. In these instances, the sun in \crd{scene a)} and the trees in \crd{scene b)} were left unchanged while other attributes \crd{of} the scenes were modified.
%It is important to note that as it completely regenerates the entire scene at each conversation turns, we show in these scenes that it is able to attend to the corresponding objects in the previous turn and copy over the relevant objects' attributes when the objects are not mentioned in the text instructions, such as the sun in scene a) and the trees in scene b) that largely remain in their original positions, while making additions and modifications to other parts of the scene.

\subsection{Sketches with \crd{Clip Art Objects} as Mask and Ratio Guidance}
The Object Generators are designed to generate sketches which respect high-level scene \crd{layout} information under the guidance of the mask and \crd{aspect} ratio conditions.
%We designed the Object Generators to generate sketches that are coherent for composing into specific sketched scenes. In this section, we demonstrate the Object Generators' performance under the guidance of mask and ratio conditions.
%\eldon{I think we can safely get rid of this next sentence. FH: Yes, I did not know why this sentence was here and I highlighted it as well.}
%As the original CoDraw task involves the placement of clip art objects by human users, our Composition Proposer was trained on their respective varying aspect ratios and poses that correspond to text modification instructions.
% The objects generated by the Composition Proposer can be represented by clip art as it learns to compose scenes from the original CoDraw task that involve the placement of clipart objects by humans users.
% These clipart objects consist of varying aspect ratios and object poses, which dictate coherant appearances of the objects that are coherent to the text modifications.
To inform generated object sketches \crd{with pose suggestions from scene composition layouts}, we built outline masks from clip art \crd{objects} and computed aspect ratios using the same method as building them for training sketches described in Section \ref{sec:object-sketchers}.
We demonstrate the Object Generator's performance in two important scenarios that allow \systemname to adapt to specific pose and subclass contexts.

\subsubsection{Generating objects for closely related classes}
\begin{figure}[h]
  \centering
  \includegraphics[width=\linewidth]{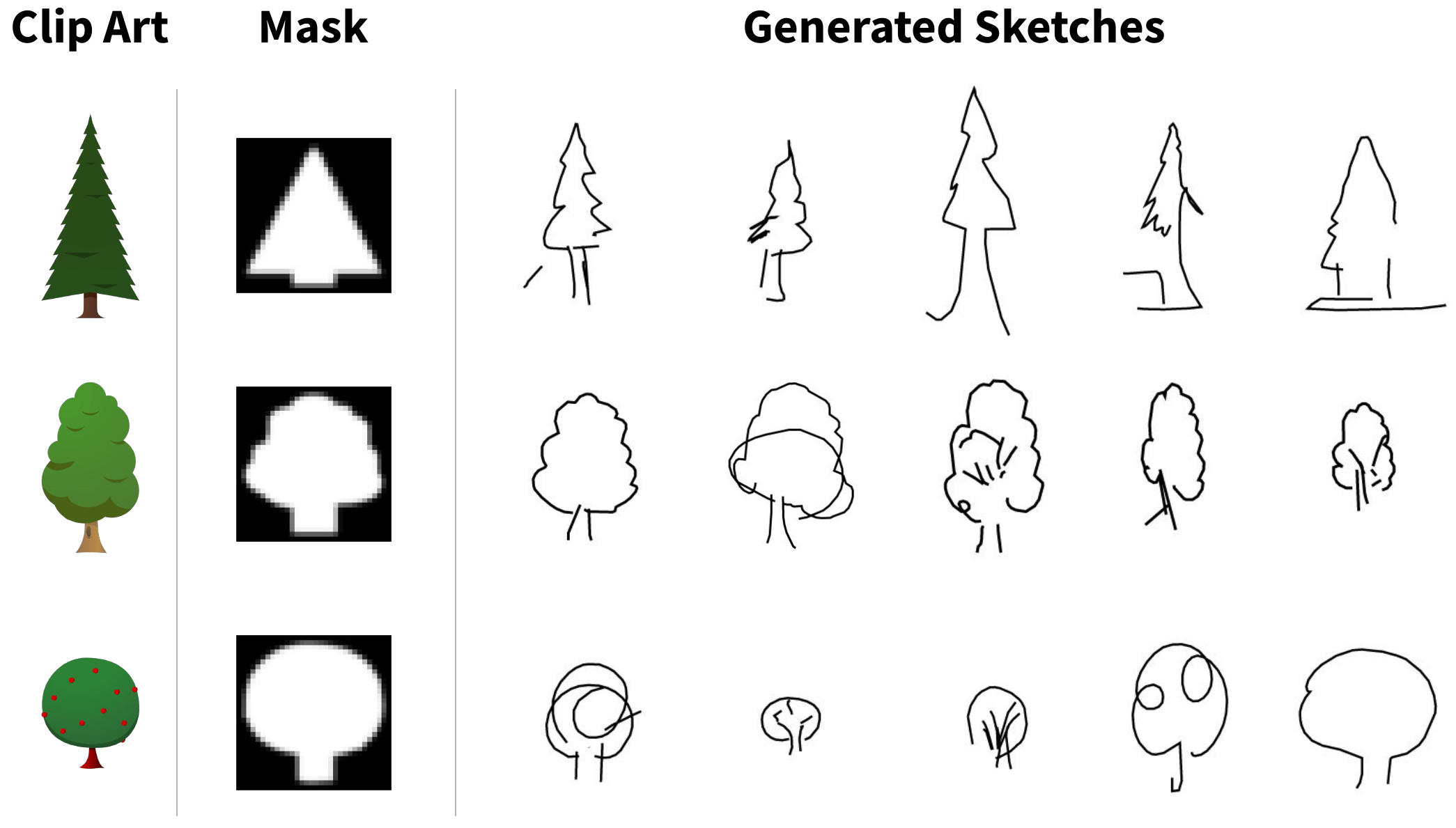}
  \caption{Sketch Generation Results \crd{of Trees} Conditioned on Masks. The Object Generator was able to sketch trees \crd{of} three different classes based on mask and aspect ratio inputs.}
  \Description[This figure shows examples of multiple kinds of trees generated by the Object Generator.]{This figure shows examples of multiple kinds of trees generated by the Object Generator. In each row, from left to right we show the clip art used to guide the generation, the mask that is generated from the clip art, and the generated sketch. On the top row, the clip art shows a cartoon-like pointy pine tree. The mask generated is largely triangular. And the generated sketches show various stroke-based trees with triangular leaves. The second row has a clip art of a cartoon-like oak tree, the mask appears to be an oval shape with curvy edges, and rectangular trunk sticking out from the bottom. The generated sketches also exhibit curved strokes for the leaves of the trees, and a solid trunk in the middle at the bottom. The final row is an apple tree with round leaves, with a mask that looks like a circle with a rectangular trunk sticking out from the bottom. The resultant sketches all demonstrate similar characteristics, with a circular stroke depicting the leaves of the tree, closing with straight trunks at the bottom.}
  \label{fig:sketch-tree}
\end{figure}

While the Composition Proposer classifies objects as one distinct class out of 58, some of these classes are closely related and are not differentiated by the Object Generators. In these cases, object masks can be used by \crd{an} Object Generator to effectively disambiguate the desired output subclass.
For instance, the Composition Proposer generates trees as one of three classes: Oak tree (tall and with curly edges), Apple tree (round and short), and Pine tree (tall and pointy); while there is only a single Object Generator trained on a general class of \crd{all types of} tree objects. We generated three different masks and aspect ratios based on three clip art images and used them as inputs to a single tree-based Object Generator to generate appropriate tree objects (by sampling $z \sim N(0, 1)^{128}$).
The Object Generator was able to sketch trees with configurations corresponding to input masks from clip art \crd{objects} (Figure~\ref{fig:sketch-tree}). The generated sketches for pine trees were pointy; for apple trees, had round leaves; and for oak trees, had curvy edges.

\subsubsection{Generating objects with direction-specific poses}
\begin{figure}[h]
  \centering
  \includegraphics[width=\linewidth]{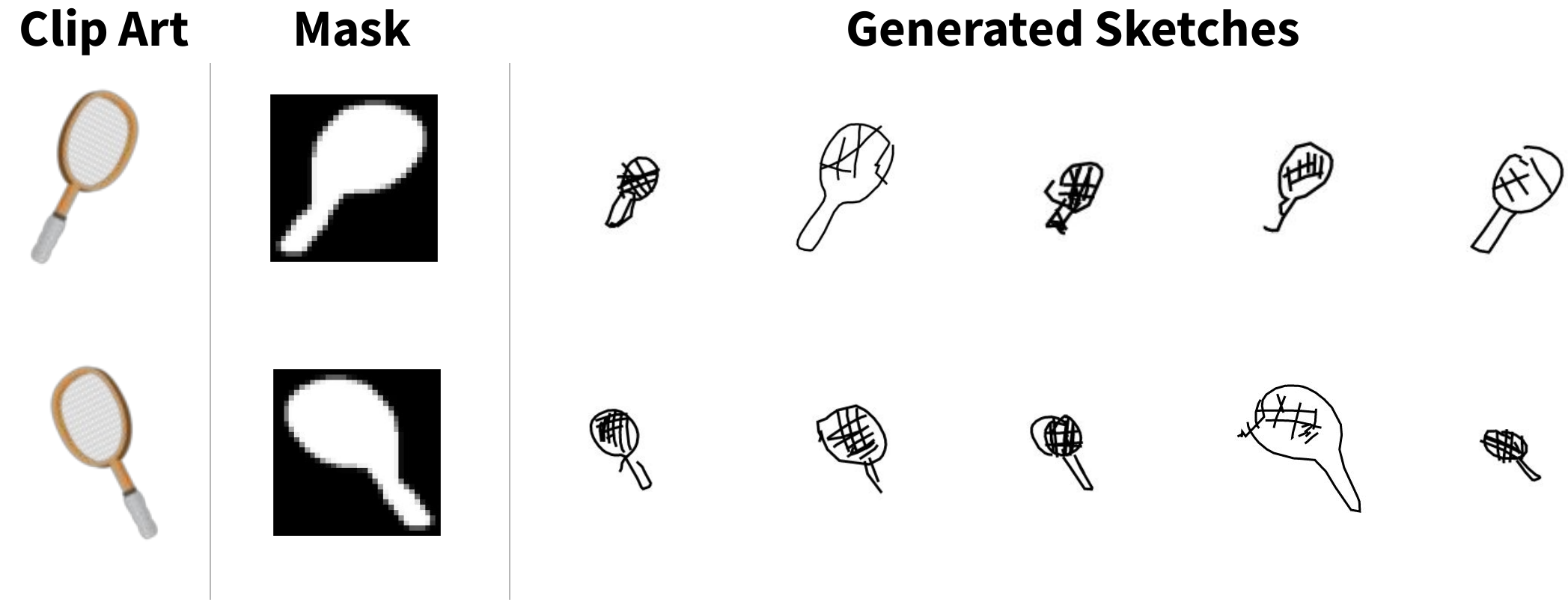}
  \caption{Sketch Generation Results \crd{of Racquets} Conditioned on Masks. The Object Generator was able to sketch racquets at two orientations consistent to the masks.}
  \Description[This figure shows examples of racquets of various orientations generated by the Object Generator.]{This figure shows examples of racquets generated by the Object Generator. In each row, from left to right we show the clip art used to guide the generation, the mask that is generated from the clip art, and the generated sketch. On the top row, the clip art shows a cartoon-like racquet with head tilting to the right, and the bottom row shows the same racquet point to the left. The mask generated is oval-shaped for the head of the racquet, and rectangular for the handle of the racquet, tilting right for the first row, and left for the second row. The generated sketches follow this pattern with sketch strokes of racquets facing right in the top row, and left in the bottom row.}
  \label{fig:sketch-racquet}
\end{figure}

The Composition Proposer can specify the horizontal orientation of the objects (pointing left or right). As such, the Object Generators are required to sketch horizontally asymmetric objects (e.g., racquets, airplanes) with a specific pose to follow user\crd{s'} instructions. We show the ability of Object Generators to produce racquets at various orientations in Figure \ref{fig:sketch-racquet}.
The generated racquet sketches conformed to the orientation of the mask, facing the specified direction at similar angles.
% The Object Generator is able to position the generated racquets with the racquet head at the top and facing the orientation as specified by the mask, and tilted at a similar angle for each orientation.

\label{sec:complete}
\subsection{Complete Sessions with Composition Layouts and Sketches}
\begin{figure}[htp]
  \centering
  \includegraphics[height=0.93\textheight, keepaspectratio]{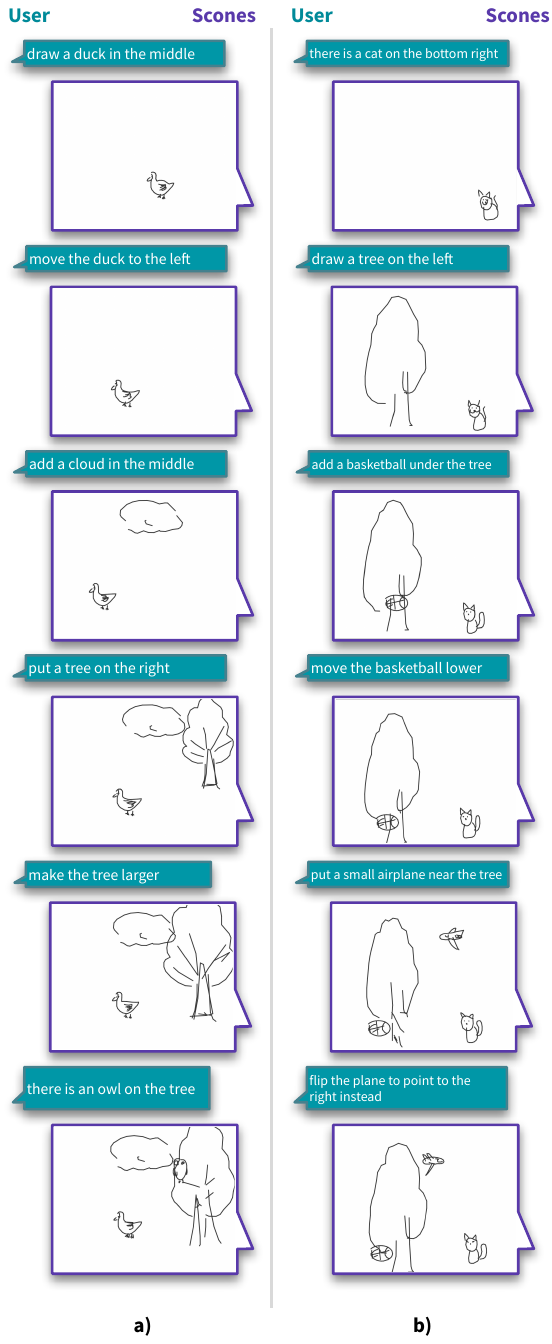}
  \caption{Complete Sketching Sessions with \systemname \crd{ curated by the authors.}
  %Each scenario illustrates modification instructions interpreted by \systemnamenospace: a) adding, translating, and scaling; and b) adding with conditions and flipping.
  }
  \Description[Two columns of example interactions with \systemname, where the user specifies instructions with text and \systemname responds by trying to draw new objects or modify existing ones.]{Left Column:
  User: draw a duck in the middle.
  System draws duck in canvas in center, slightly to the right.
  User: move the duck to the left.
  System moves the duck slightly left of center.
  User: add a cloud in the middle.
  System draws cloud in top center. Duck moves slightly left.
  User: put a tree on the right.
  System draws a tree to the top right of the canvas. Bottom of tree is just below center line of canvas.
  User: make the tree larger.
  System enlarges tree proportionally, trunk now almost reaches the bottom and the tree takes a third of the canvas roughly.
  User: there is an owl on the tree.
  System draws an owl perched on a branch drawn as a line in the tree.
  Right column:
  User: there is a cat on the bottom right.
  System draws a cat on bottom right.
  User: draw a tree on the left.
  System draws a tree in the left half of the canvas.
  User: add a basketball under the tree.
  System draws a basketball in the center of the tree.
  User: move the basketball lower.
  System redraws basketball slightly lower and to the left.
  User: put a small airplane near the tree.
  System draws a small airplane in the top right, facing left.
  User: flip the plane to point to the right instead.
  System redraws the airplane closer to the tree, now pointing right.}
  
  \label{fig:complete}
\end{figure}

% Combining the Composition Proposer and Object Generators, we qualitatively evaluate the performance of \systemname holistically by inspecting generated scenes given input from the test set of the CoDraw dataset.
We show the usage of \systemname in six turns of conversation from multiple sessions in Figure~\ref{fig:complete} and Figure~\ref{fig:teaser}. \crd{We curated these sessions by interacting with the system ourselves to demonstrate various capabilities of \systemnamenospace.} In \crd{session a)}, \systemname was able to draw and move the duck to the left, sketch a cloud in the middle, and place and enlarge the tree on the right, following instructions issued by the user. In \crd{session b)}, \systemname was similarly able to place and move \crd{a cat, a tree, a basketball and an airplane}, but at different positions from \crd{session a)}.
For instance, the tree was placed on the left as opposed to the right, and the basketball was moved to the bottom. We also show the ability of \systemname to flip objects horizontally in \crd{session b)}, such that the plane was flipped horizontally and regenerated given the instructions of ``flip the plane to point to the right instead''. This flipping action demonstrates the Object Generator's ability to generate \crd{objects with the require poses} by only sharing the latent vectors $z$, such that the flipped airplane exhibits similar characteristics as the original airplane. In both sessions, \systemname was able to correlate multiple scene objects, such as placing the owl on the tree, and basketball under the tree in \crd{session b)}.

Moreover, we discover that \systemname was able to handle more advanced instructions, such as generating multiple objects at once. In Figure~\ref{fig:teaser}, other than basic enlarging and moving commands for the pizza and \crd{the} helicopter, \systemname was able to sketch both the sun and the cloud onto the scene, at positions satisfying the first instruction. It was also able to sketch a table directly under the pizza with the instruction `put the pizza on a table'. 

%These scenes demonstrates Scones' competitive performance producing coherent objects relevant to multiple types of instructions issued by human users.

\subsection{Interpreting Transformer's Attention Maps}
\label{sec:attn}
We can further verify the relationship between text and object representations learned by the model by visualizing attention weights computed by the Transformer model of the Composition Proposer. These weights \crd{also} create the unique possibility of generalizing and prompting for sketches of new objects specified by users.

\begin{figure}[h]
  \centering
  \includegraphics[width=\linewidth]{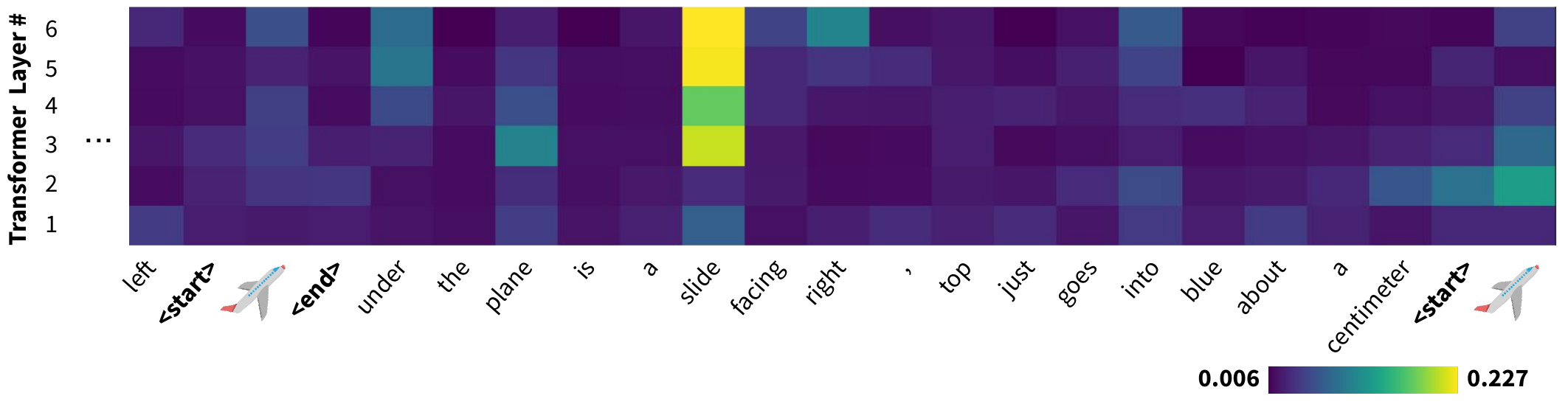}
  \caption{Attention Map of the Transformer across Object and Text Tokens for the Generation of Slide in the Second Turn of Conversation. We observed that the Transformer model attended to the corresponding words and objects that describe objects in the scene \crd{related to the newly generated `slide' object.}}
  \Description[A 2D heatmap showing the word ``slide'' is most attended to by the transformer.]{A 2D heatmap with input tokens (text and previous object sketches from the scene) on the horizontal axis (``[\dots] <start> [airplane sketch] <end> under the plane is a slide facing right, top just goes into blue about a centimeter <start> [airplane sketch]'') and Transformer network layers 1-6 on the vertical axis. The attentions to the word ``slide'' are prominent, and with other inputs such as ``under'',  ``plane'', and the second airplane sketch being slightly activated.}
  \label{fig:attn-second}
\end{figure}

%The Transformer model uses masked self-attention to attend to the outputs of previous time steps in the conversation.
The Transformer model in the Composition Proposer uses masked self-attention to attend to scene objects and instructions from previous time steps most relevant to generating an object specification at the current turn.
% In the context of sketch critique, it attends to previous scene objects or words in the critique text that's most relevant to the generation of scene objects in each time-step.
We explore the attention weights of the first two turns of a conversation from the CoDraw validation set.
In the first turn, the user instructed the system, ``top left is an airplane medium size pointing left''.
When the model generated the first object, it attended \crd{to} the ``airplane'' and ``medium'' text tokens to select class and output size.
In the second turn, the user instructed the model to place a slide facing right under the plane.
The model similarly attended to the ``slide'' token the most, while also \crd{significantly} attended to the ``under'', and ``plane'' text tokens\crd{, and the plane object}, which are useful for situating the slide object at the desired location relative to an existing airplane object (Figure~\ref{fig:attn-second}).

\begin{figure}[h]
  \centering
  \includegraphics[width=0.7\linewidth]{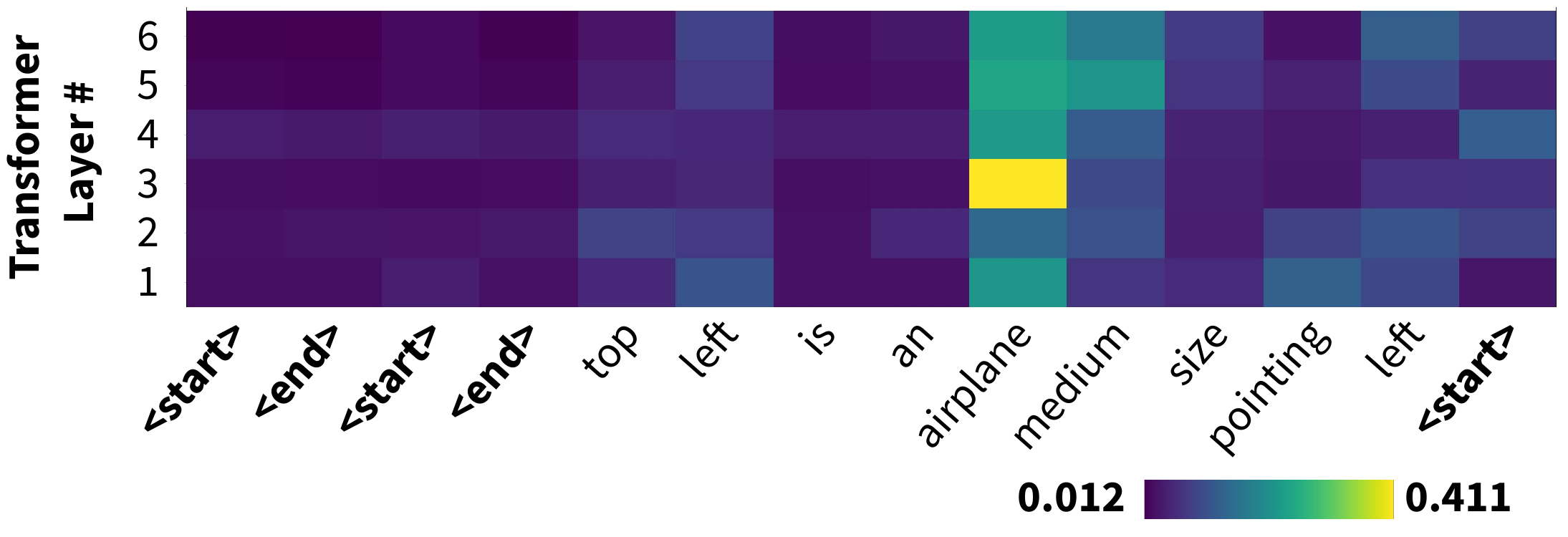}
  \caption{Attention Map of the Transformer across Object and Text Tokens for the Generation of an Airplane, the First Object in the Scene.}
  \Description[A 2D heatmap showing the word ``airplane'' is the most attended by the Transformer.]{A 2D heatmap with input tokens from a turn on the horizontal axis (``<start> <end> <start> <end> top left is an airplane medium size pointing left <start>'') and Transformer network layers 1-6 on the vertical axis. The attentions to the word ``airplane'' are particularly prominent, and ``medium'' is slightly prominent as well.}
  \label{fig:attn-first}
\end{figure}

%As shown in Figure \ref{fig:attn-second}, when generating the new slide object in the scene, the model similarly attends to the `slide' text token the most, while also attending to `under', `plane' text tokens, and also the `plane' object which are all useful for generating the slide object at a coherent location related to the plane, an existing object in the scene.

\begin{figure}[h]
  \centering
  \includegraphics[width=\linewidth]{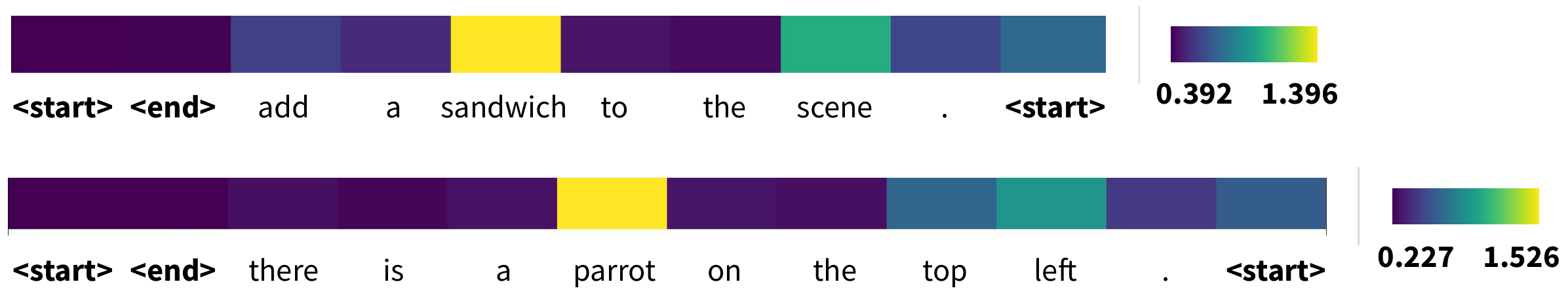}
  \caption{Attention Map of the Transformer for Text
  \crd{Instructions} that Specifies Unseen Objects.}
   \Description[1D heatmaps showing the level of attention for each text token for instructions with unseen objects, highlighting ``sandwich'' and ``parrot''.]{1D heatmaps with text tokens on the horizontal axes (``add a sandwich to the scene . <start>" and ``there is a parrot on the top left . <start>'') and aggregated attention values as color blocks for each token along the axes. The attentions to the words ``sandwich'' and ``parrot'' are prominent.}
  \label{fig:attn-ooc}
\end{figure}

These attention weights could be used to handle unknown scene objects encountered in instructions.
% especially at the first turn of conversation when we can make the assumption that the users intend to add an object.
When the model does not output any scene objects, but only a $o_e$ (scene end) token, we can inspect the attention weights for generating this token to identify a potentially unknown object class, and ask the user for clarification. For example, when users request unsupported classes, such as a `sandwich' or  `parrot' (Figure \ref{fig:attn-ooc}), \crd{\systemname could identify this unknown object by taking the text token with the highest attention weight, and prompting the user to sketch it by name.}
%\systemname could share \crd{its knowledge of the potential class of the object (by taking the text token with the highest attention weight) to keep users engaged, while prompting them to complete the sketch of the unknown class.}
%\eldon{Instead of `keeping users engaged', \systemname could identify this unknown object by taking the test token with the highest attention weight, and prompting the user to sketch it specifically by name. FH: Addressed}
% When the model infers a user wishes to add an object, but encounters an unknown 
%When the model was not able to output novel scene objects but a single $o_e$ (scene end) token (as expected) in the first turn, we can inspect the attention weights on the caption and consider the word in the previous turn of text that receives most attention weight as the name of the object that user desires to sketch.

%In Figure \ref{fig:attn-ooc}, we show examples when the user is trying to add a sandwich or a parrot to the scene, which are both classes that are not currently supported by the Composition Proposer. From the attention weights of the model when generating the `scene end' token, we observe the model was able to attend to the word `sandwich' and `parrot', which is sufficient to provide prompts for user to collaborate with the system such as \textbf{`I am not sure how to draw \emph{<the word with the most attention>}, can you help me out?'}. This provides a collaborative environment and let users perceive the system to be partially understanding the scene, even when the object is out of the original class vocabulary. 

\section{Exploratory Evaluation}
\label{sec:user-study}
To determine how effectively \systemname can assist users in creating sketches from natural language, we conducted an exploratory evaluation of \systemnamenospace. We recruited 50 participants from English-speaking countries on Amazon Mechanical Turk (AMT) for our study. We collected quantitative and qualitative results from user trials with \systemnamenospace, as well as suggestions for improving \systemnamenospace. Participants were given a maximum of 20 minutes to complete the study and were compensated \$3.00 USD. Participants were only allowed to complete the task once.

\subsection{Method}
\crd{Participants asked to recreate one of five randomly chosen target scene sketches by providing text instructions to \systemname in the chat window.}
Each target scene had between four and five target objects from a set of 17 scene objects. Participants were informed that the final result did not have to be pixel perfect to the target scene, and to mark the sketch as complete once they were happy with the result. Instructions supplied in the chat window were limited to 500 characters, and submitting \crd{an instruction} was considered as taking a ``turn''. The participants were only given the sketch strokes of the target scene without class labels, to elicit natural instructions. 

\begin{figure}[h]
  \centering
  \includegraphics[width=\linewidth]{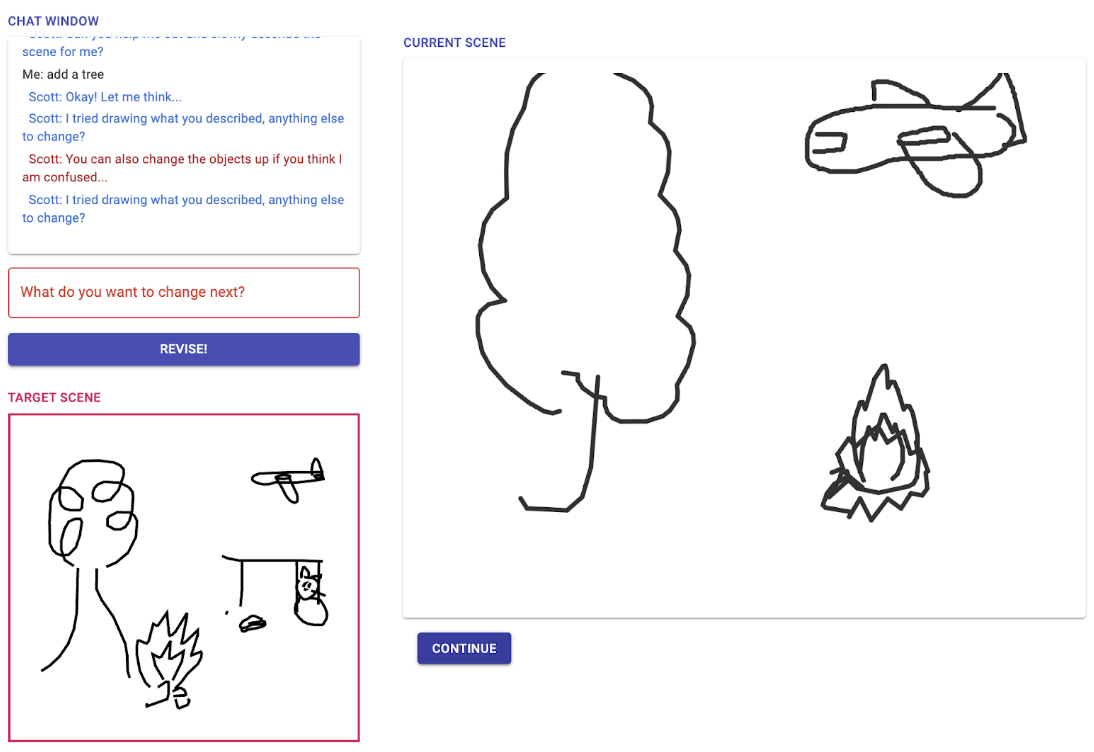}
  \caption{Screenshot of \systemnamenospace's Evaluation User Interface.}
  \Description[Web application with chat window, target scene, and current scene.]{A web application with a chat window to the top left, a ``target scene'' window on the bottom left, and a ``current scene'' pane to the right. Target scene has a tree on the left, a campfire to the bottom center; and an airplane, desk, cat, and sandwich to the right. Current scene has a tree on the left, and airplane on the top right, and a campfire on the bottom right.}
  \label{fig:study-interface}
\end{figure}

Participants were first shown a short tutorial describing the canvas, chat interface, and target scene in the \systemname interface (Figure~\ref{fig:study-interface}), and were asked to give simple instructions in the chat window to recreate the target scene.
\crd{Only two sample instructions were given in the background image of the tutorial: ``add a tree'', and ``add a cat next to the table''.} 
At each turn, participants were given the option to redraw objects which remained in the scene for over three turns using a paintbrush\crd{-based} interface. After completing the sketch, participants filled out an exit survey with likert-scale questions on their satisfaction at the sketch and enjoyment of the system, \crd{and open-ended feedback on the system.} %Questions included open-ended responses and Likert scale questions (1 = ``Strongly Disagree'', 5 = ``Strongly Agree''). Likert scale questions were phrased as follows: \textit{(Satisfaction): ``I am satisfied with the final sketch I produced together with the system''; (Enjoyment): ``I enjoyed the conversational sketching task''.}

\subsection{Results}

\subsubsection{Participants Satisfied with Sketches, Enjoyment Was Bimodal}
Participants were generally satisfied with their final sketches ($\mu = 3.38, \, \sigma = 1.18$), and enjoyed the task ($\mu = 4.0, \, \sigma = 1.12$).
In open-ended feedback, participants praised \systemnamenospace's ability to parse their instructions: \textit{``it was able to similarly recreate the image with commands that I typed'' (P25); ``I liked that it would draw what I said. it was simple and fun to use'' (P40).}
Some participants even felt \systemname was able to \emph{intuitively} understand their instructions. P15 remarked, \textit{``I thought it was cool how quickly and intuitively it responded,''} while P35 said, \textit{``It had an intuitive sense of what to draw, and I did not feel constrained in the language I used''.}
%Many participants were additionally impressed by \systemnamenospace's speed: \textit{``quick response'' (P32), ``I enjoyed how quick it was'' (P2),} and \textit{``[I liked] How fast the system responded'' (P22).}

\begin{figure}[h]
  \centering
  \includegraphics[width=\linewidth]{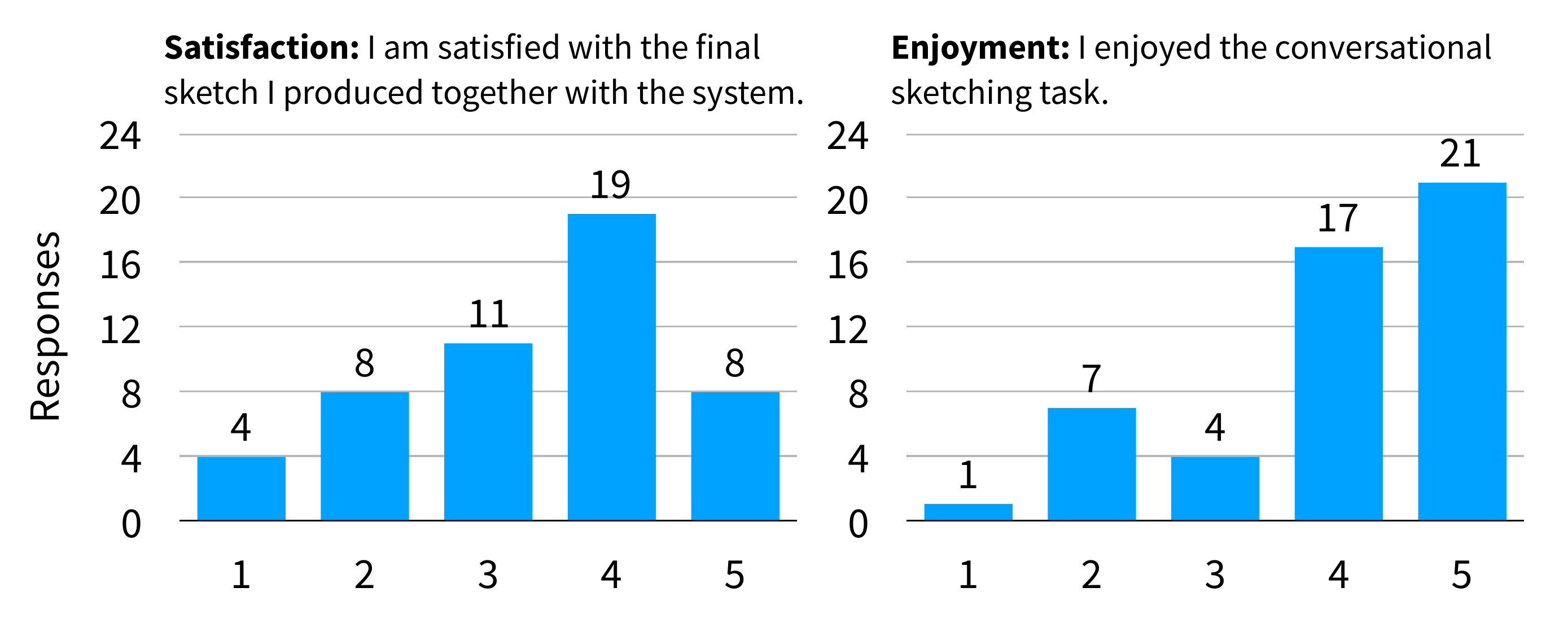}
  \caption{Survey Results from User Sessions with Scones.}
  \Description[Two histograms showing participant satisfaction and enjoyment from likert scale questions.]{Left: histogram of agreement to ``I am satisifed with the final sketch I produced together with the system''. 4 strongly disagree, 8 disagree, 11 neutral, 19 agree, 8 strongly agree.
  Right: histogram of agreement to ``I enjoyed the conversational sketching task''. 1 strongly disagrees, 7 disagree, 4 neutral, 17 agree, 21 strongly agree.}
  \label{fig:likerts}
\end{figure}

While enjoyment was high on average, we found responses to enjoyment followed a bimodal distribution (Figure~\ref{fig:likerts}). By reviewing qualitative feedback and instructions to \systemnamenospace, we observe that many instances of low enjoyment (score $\leq 2$) come from class confusion in the target scene sketch. Some participants confused the tent in a target scene as a ``pyramid'' in their instructions, which \systemname does not support: \textit{``There is a pyramid on the left side a little ways up from the bottom'' (P44).} P49 tried five times to add a ``pyramid'' to the scene.
%Excerpts from their instructions include \textit{``There's a pyramid under the airplane'', ``Put a pyramid on the bottom left'', ``put a pyramid on the bottom left, \dots''}.

P17, who strongly disagreed with enjoying the task (1/5), faced repeated class confusion issues, mentioning, \textit{``it was very frustrating that it wouldn't draw the circle by the cloud \dots It wouldn't draw anything besides the plane, cloud, tent, and fire. Was that not a person up by the cloud?''} \systemname does not support ``circle'' or ``person'' classes---the target sketch had the sun next to the cloud. When \systemname is asked to draw an unsupported object, the canvas will be left unchanged. Providing participants \crd{with} an explicit list of classes in the target image or adding error messages could mitigate these frustrations. Furthermore, attention-based methods mentioned in Section \ref{sec:attn} could be used when an unrecognized class is detected to prompt users to provide sketch strokes with a corresponding label.

\subsubsection{Participants Communicate with \systemname at Varying Concept Abstraction Levels}
\begin{figure}[h]
  \centering
  \includegraphics[width=\linewidth]{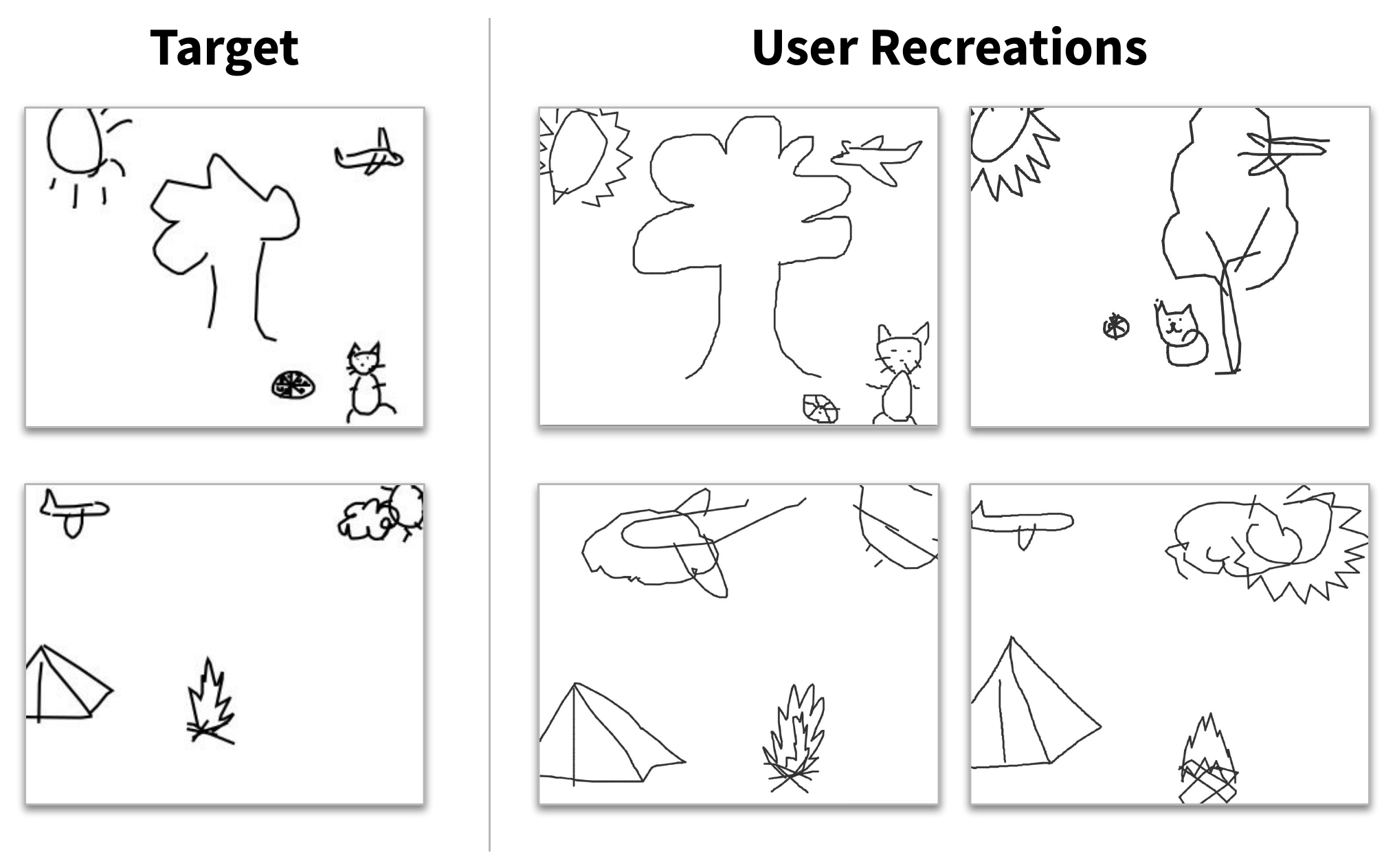}
  \caption{Recreated Scenes during the User Study. Users combined \systemnamenospace-generated outputs with their own sketch strokes to reproduce the target scenes presented to them.}
  \label{fig:recreations}
\end{figure}

On average, participants completed the sketching task in under 8 turns ($\mu = 7.56, \, \sigma = 3.42$), with a varied number of tokens (words \crd{in instructions}) per turn ($\mu = 7.66, \, \sigma = 3.35$). Several participants only asked for the objects themselves (turns delimited by commas): \textit{``helicopter, cloud, swing, add basketball'' (P25)}.
Other participants made highly detailed requests: \textit{``There is a sun in the top left, There is an airplane flying to the right in the top right corner, There is a cat standing on it's hind legs in the bottom right corner, Move the cat a little to the right, please, \dots'' (P14).}
Participants who gave instructions at the expected high-level detail produced satisfying results, \textit{``draw a tree in the middle, Draw a sun in the top left corner, A plane in the top right, A cat with a pizza under the tree'' (P32).}  The recreation of this participant is shown on the top right of Figure~\ref{fig:recreations}.

The longest conversations were often from participants with mismatched expectations for \systemnamenospace, who repeated commands: \textit{``Draw a cloud in the upper left corner with three round edges., Change the cloud to have 3 round edges., Draw only 3 round waves around the edge of the cloud., \dots Draw a snowman to the left of the table., \dots Draw a circle touching the middle circle., \dots'' (P23).} This trial reflects the need for \systemname to make clearer expectations of input to users. \crd{P23's 16-instruction session contains} expectations for the system to modify low-level aspects of the sketches (changing the number of edges in the cloud), exhibits class confusion (snowman and circles with shovel), and has mismatched concept abstraction levels (drawing a shovel versus constructing a shovel from visual primitives, i.e., circles). A potentially simple mitigation for these hurdles would be to introduce more detailed tutorial content for a wider deployment of \systemnamenospace.

\subsubsection{\systemname as a Tool for Collecting Iterative Sketching Data}
The results of our study show significant potential for \systemname to be used as a Game With a Purpose (GWAP)~\cite{gwap} to collect sketch critiques (natural language specified modifications to an input \crd{sketch} to match a target sketch) and user-generated sketching strokes. 26 ($52\%$ of) participants redrew objects in their sketches when prompted ($\mu = 0.98, \, \sigma = 1.19$), and participants who redrew objects expressed their appreciation for this feature: \textit{``I liked that I could redraw the image'' (P48); ``I liked being able to draw parts myself because it was relaxing and I felt I was more accurate'' (P11).} Most participants who redrew objects also kept output from \systemname in their final sketches, reflecting \systemnamenospace's potential as a mixed-initiative design tool. Redrawing was voluntary in our task, and these results suggest \systemname may be useful for collecting user-generated sketches in addition to natural language critique in a GWAP. Further motivating this application, 14 participants described the task as ``fun'' in open-ended feedback, e.g., \textit{``This was a very fun task'' (P23); ``(I liked) Playing the game and describing the drawing. It was fun!'' (P42).}

% \todo{redrawing stats. Examples of redrawn things (need a fig). Comments from users who redrew things -- they liked it!}
% \todo{Fun quotes: 
% P5
% fun and I had a good time trying to get the picture remade
% P12
% That was fun. Hopefully I did this correctly. The computer did well.
% P13
% Very fun; would be interesting to give some complex examples of dialogue that Scott can interpret.
% P16
% Was fun and different and also I'm an awful artist.
% P19
% this was fun
% P20
% Nothing I can think of. Had a lot of fun with this - thanks! :)
% P23
% Good luck with your research! This was a very fun task.
% P26
% It was fun to see how it would respond to my commands.
% P31
% I really enjoyed it, it was very fun.
% P36
% It was fun to see how the system interpreted my directions and enjoyed describing the scene.
% P40
% I liked that it would draw what I said. it was simple and fun to use.
% P42
% Playing the game and describing the drawing. It was fun!	
% P43
% It seemed fun but could use some tweaking.
% P48
% Not really, I don't know exactly what is being studied. It was fun and interesting, thank you

% Similar:
% P33
% I thought that ift was pretty interesting and quite engaging to take part in this task.
% P8
% This study was very enjoyable, the system is unique.

% }
% \todo{People said it was fun. They redrew 1 thing on average. Some users kept in generated sketches too. People enjoyed redrawing.}

\subsection{Participants' Feedback for Improving \systemnamenospace}
Participants offered suggestions for how they would improve \systemnamenospace, providing avenues for future work.

\subsubsection{\crd{Object} Translations and Spatial Relationships}
A major theme of dissatisfaction came from the limited ability of our system to respond to spatial relationships and \crd{translation-related} instructions at times: 
\textit{``It does not appear to understand spatial relationships that well'' (P35); ``you are not able to use directional commands very easily'' (P11).}
These situations largely originate from the CoDraw dataset~\cite{codraw}, in which users had a restricted view of the canvas, resulting in limited relative spatial instructions. This limitation is discussed further in Section \ref{sec:data-lim}.

To improve the usability of \systemnamenospace, participants suggest its interface could benefit from \crd{the addition of} direct manipulation features, such as selecting and manually transforming objects in the scene: \textit{``I think that I would maybe change how different items are selected in order to change of modify an object in the picture. (P33); ``ma\crd{yb}e there should be a move function, where we keep the drawing the same but move it'' (P40).}
%``Everything was going perfect until I asked it to draw a ball. After that, it started drawing a bunch of lines over the swing instead. I thought i would be able to get it redraw objects after I tried to manually add the ball, but everything got kind of screwed up after that'' (P3). ''}
Moreover, some participants also recommended adding an undo feature, \textit{``Maybe a separate button to get back'' (P31)}, or the ability to manually invoke \systemname to redraw an object, \textit{``I'd like a way to ask the computer to redraw a specific object'' (P3)}. These features could help participants express corrective feedback to \systemnamenospace, potentially creating sketches that better match their intent.

\subsubsection{More Communicative Output}
Some participants expected \systemname to provide natural language output and feedback to their instructions.
Some participants asked questions directly to elicit \systemnamenospace's capabilities: \textit{``In the foreground is a table, with a salad bowl and a jug of what may be lemonade. In the upper-left is a roughly-sketched sun. Drifting down from the top-center is a box, tethered to a parachute., Did you need me to feed you smaller sentences? \dots'' (P38)}.
P23 explicitly suggested users should be able to ask \systemname questions to refine their intentions: \textit{``I would like the system to ask more questions if it does not understand or if I asked for several revisions. I feel that could help narrow down what I am asking to be drawn''.}
Other participants used praise between their sketching instructions, which could be used as a cue to preserve the sketch output and guide further iteration: \textit{``\dots Draw an airplane, Good try, Draw a table \dots'' (P1); ``Draw a sun in the upper left corner, The sun looks good! Can you draw a hot air balloon in the middle of the page, near the top? \dots'' (P15).}
% ``draw helicopter on the top left, Great! Draw a cloud on the top right \dots'' (P12);
Providing additional natural language output and prompts from \systemname could enable users to refine \systemnamenospace's understanding of their intent and learn about system capabilities. A truly \emph{conversational} interface with a sketching support tool could pave the way for advanced mixed-initiative collaborative design tools.

\section{Limitations}
\label{sec:limit}
\subsection{Underspecified Masks}
\begin{figure}[h]
  \centering
  \includegraphics[width=\linewidth]{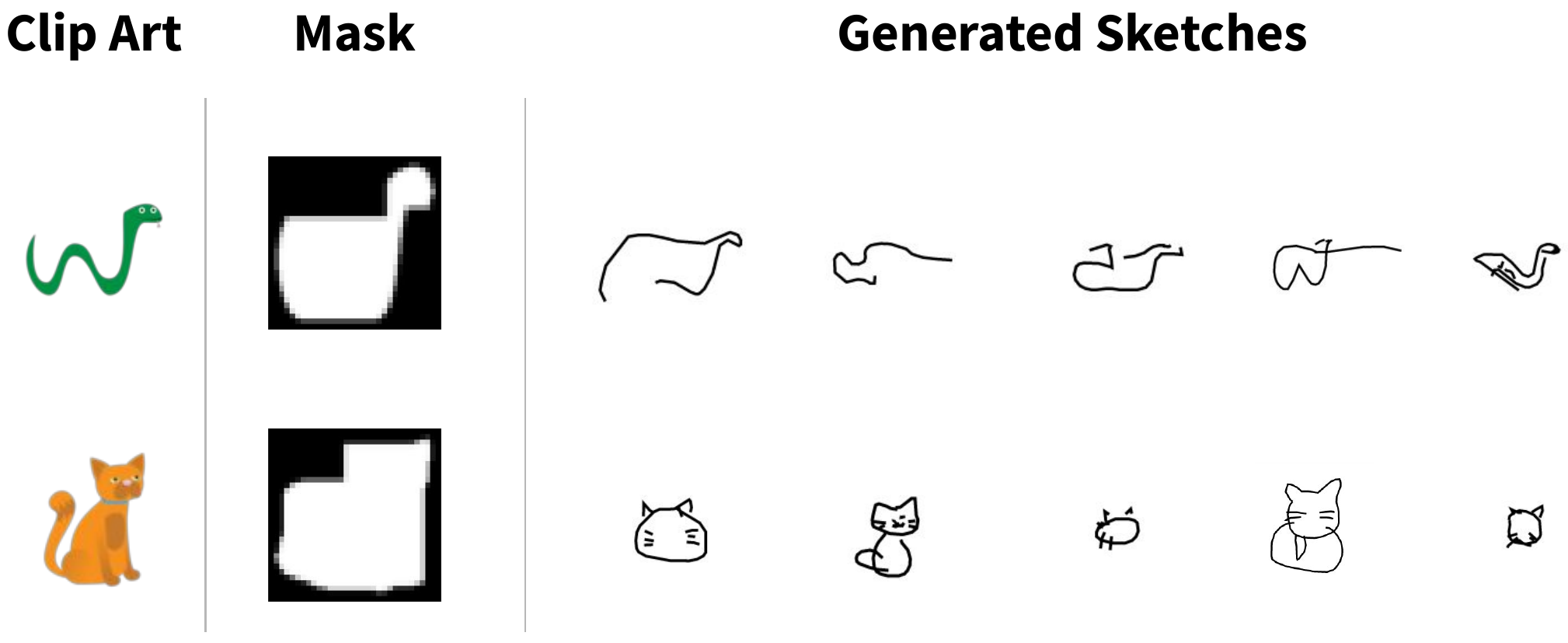}
  \caption{Sketches Generated by the Object Generator with Underspecified Masks of the Snake and Cat Classes.}
  \Description[This figure shows clip art objects, masks, and generated sketches by the Object Generator with Underspecified Masks of the snake and cat classes.]{This figure shows clip art objects, masks, and generated sketches by the Object Generator with Underspecified Masks of the snake and cat classes. In the top row, we show the clip art with a wavy snake. However, this snake outline is concave, and hence the mask only captures two rectangular blocks. This results in sketches generated with shapes that are rectangular, but doesn't follow the original clip art object. Similarly, the cat clip art object is a squatting cat, which converts to a rectangular mask. However, the face of the cat also translates to a rectangular mask, resulting in sketch strokes depicting faces of cats generated by the Object Generator.}
  \label{fig:limitations}
\end{figure}

While mask conditioning effectively guides the Object Generator\crd{s} in creating sketches with desired configurations, they can be underspecified for the poses exhibited by objects of some classes.
% Despite Scones' improvements in resolving the ambiguity in the poses of generated objects over existing systems by incorporating masks of the objects in the Object Generator, the aliasing of poses can still occur for various classes.
As shown in Figure \ref{fig:limitations}, the mask of a right-facing body of a sitting cat can be similar to the face of a cat. The current mask generation algorithm is also not able to capture all the curves of the snake, resulting in ambiguous sketches of snakes. \crd{Future iterations of \systemname can improve on the mask generation algorithms with more advanced techniques.}

%\sig{In general, our current mask generation algorithm can only partially capture the shape of most concave objects.} 
%in poses that can impact the quality of the sketches.
%Taking a majority vote over multiple lines at multiple angles in the mask generation algorithm, instead of \crd{using merely} horizontal and vertical lines at any given point, could better capture the outline of the sketches and clip art \crd{objects}.

% Limitations from the dataset
\subsection{Limited Variation of Sketches}
\systemname currently supports a limited number of sketched object classes and poses due to its discrete representation of object configurations used by the Composition Proposer. 
%This limitation originates from the CoDraw dataset that it derives conversational critique knowledge from.
Future work should explore models conditioned on continuous representations of classes and poses from word embeddings for a flexible number of object classes. Moreover, \systemname currently supports only limited stylistic modifications (i.e., it may not support `sketch the leaves on the tree with more details'). A future iteration of the Composition Proposer could output a continuous embedding that contains objects' class, pose, and stylistic information to fully support a wide range of sketches.  
%Future work can explore developing a general-purpose Object Generator for all object classes that is conditioned on the word embedding of the class names of the generated sketch objects.

%Moreover, as mask and aspect ratio conditions used by the Object Skethcers already offer the capability of sketching objects with a large variety of poses, we can replace the discrete representation of object sizes and masks in the Composition Proposer and adopt to these new object sizes using online-training while Scones is deployed as a game. 

\subsection{Data Mismatch Between CoDraw and Target Task}
\label{sec:data-lim}
There are differences between the task protocol used to collect \crd{the} CoDraw \crd{dataset} and the user interactions in \systemnamenospace. The conversation in CoDraw only offers the Teller one chance to `peek' at the Drawer's canvas, which significantly decreases the number of modifications to existing scene objects. As a result, \systemname performs well at adding \crd{objects of} correct classes at appropriate sizes, but is not as advanced at modifying or removing objects. Future work can explore data augmentation techniques, such as super-sampling randomly-perturbed rounds with modifications, or adding removal rounds that mirror the addition of scene objects, to improve the ability of \systemname to handle these tasks. 

%Extending beyond mask-based techniques, we can augment the Sketch-RNN model with text captions that users have specified in GWAP to annotate the latent space using these text captions. We can form multiple triplets of target sketch, current sketch, and critique text for various objects, and derive multiple directions in the latent space that correspond to these text captions and use them to generate relevant objects in the latent space in the future. 

\section{Discussion and Future Work}
\subsection{Scones-supported Games With a Purpose (GWAP)}
\label{sec:gwap}
While Scones \crd{demonstrates} a plausible system architecture for composing scenes of limited scenarios, the limitations of object classes and modification capability are mainly due to the lack of large-scale datasets of multimodal sketch modification. These datasets are considered to be difficult to collect due to the sketching skill requirement of crowdworkers~\cite{sketchyscenes}. We believe \systemname can be used as a gateway towards creating such a dataset. By decomposing each scene into \crd{object components}, crowdworkers would only need to sketch a single object in context, which was shown to be possible from the Quick, Draw! dataset. While the models currently are restricted to \crd{handling} objects of a small set of poses and aspect ratios, we can prompt users to generate these objects freely, in turn expanding the variety of sketches in our dataset.
% We could furthermore retrain our model to consider a wider range of objects.
Since \systemname can automate sketch generation for other parts of the scene, this significantly improves the scalability of the game and makes \crd{it} possible \crd{for \systemname to be used} as a data-collection system. Moreover, we also collect text instructions that could help to build a critique model for providing \crd{text-based sketch modification suggestions} to users.
%The intelligence component of Scones shall also improve the playability and enjoyability of the game as shown in \ref{sec:user-study}. 

%\subsection{Sketch-augmented Voice Assistants}
%Since \systemname{} supports multimodal interaction in sketching and text, we believe it has the potential to support voice assistants that is frequently used in situations where direct manual interactions are not possible by users, such as for accessibility or multi-tasking (e.g., cooking) applications, to create drawings. Moreover, we believe the types of sketches generated by \systemname{} is aligned with the conversational style text that these voice assistants generates, which can be naturally extended to allow users to generate engaging graphic novels through storytelling activities.

\subsection{Application to Professional Domains}
%\systemname currently focuses on modifying and composing general, informal scene objects.
%\eldon{informal?}
%\sig{With the improvements proposed in Section \ref{sec:limit} to support more flexible sketches with varied style, classes and poses, and the suggestions proposed in Section \ref{sec:gwap} to collect new data that contain more robust information about the sketches and text instructions,
\sig{We believe the system architecture of \systemname can be applied to professional domains if object and scene data for these domains become available.
For instance, \systemname could participate in the UI/UX design process by iteratively modifying UI design sketches according to design critique. To enable this interaction, we could consider complete UI sketches as `scenes' and UI components as `scene objects'. \systemname could be trained on this data along with text critiques of UI designs to iteratively generate and modify UI mockups from text. While datasets of UI layouts and components, such as those presented in Swire~\cite{swire}, suggest this as a near possibility, this approach may generalize to other domains as well, such as industrial design.
% and the placement and generation of these components as scene composition and object generation tasks, according to the abstractions introduced in this paper. We will then need to instead train \systemname with sketches of UI designs and UI layout data (e.g., those presented in Swire \cite{swire}), along with text critique of the generation process.
%This approach could generalize to other domains such as sketching for industrial design.
%Similarly, we can consider individual parts of products as scene objects when applying \systemname to Industrial Design. \eldon{I think we can scratch the industrial design example? How specific do we need to be about implementation details for \systemname with UX design? Can't we just say it's possible with datasets such as those from Swire?}}
Future work in adapting our system to new domains could benefit from fine\crd{-}tuning pre-trained models in \crd{the current implementation of \systemnamenospace.}}
%that already have basic knowledge of object placements, orientations, and sizes.

\section{Conclusion}
In this paper, we introduced \systemnamenospace, a machine-learning-driven system that generates scenes of sketched objects from text instructions. \systemname consists of two stages, a \emph{Composition Proposer} and a set of \emph{Object Generators}, to compose sketched scenes of multiple objects that encode semantic relationships specified by natural language instructions. \crd{We establish state-of-the-art performance for the text-based scene modification task}, and introduce mask conditioning as a novel component in the Object Generators, which enables finer-grained control of object poses in sketch output. 
With \systemnamenospace, users can interactively add and modify objects in sketches with inferred operations (e.g., transforming, moving, reflecting).

In an exploratory user evaluation, we found participants enjoyed working with \systemname and were satisfied with the output sketches it produced. Most participants contributed hand-drawn sketches during the activity, motivating the potential for \systemname to be used as a Game With A Purpose (GWAP) for collecting end-to-end sketching critique and modification data.

We see \systemname as a step towards design support interfaces with tight human-in-the-loop coupling, providing an entirely new means for creative expression and rapid ideation. We are excited to continue designing for this future of design, art, and engineering.
\begin{acks}
The authors would like to thank Douglas Eck, Heiga Zen, Yingtao Tian, Nathaniel Weinman, J.D. Zamfirescu-Pereira, Philippe Laban, David M. Chan, Roshan Rao, and all anonymous reviewers for providing valuable comments on the system and the paper. The authors would also like to thank all workers on Amazon Mechanical Turk for participating in the study and providing valuable feedback and comments on the system. This research is supported by Google Cloud.
\end{acks}

\bibliographystyle{ACM-Reference-Format}
\bibliography{ref}

\end{document}